\documentclass[pdflatex,sn-mathphys-num]{sn-jnl}

\usepackage{graphicx}%
\usepackage{multirow}%
\usepackage{amsmath,amssymb,amsfonts}%
\usepackage{amsthm}%
\usepackage{mathrsfs}%
\usepackage[title]{appendix}%
\usepackage{textcomp}%
\usepackage{manyfoot}%
\usepackage{multirow}
\usepackage{booktabs}%
\usepackage[normalem]{ulem}

\usepackage{algorithmicx}%
\usepackage{listings}%
\usepackage{subcaption}
\usepackage{braket} 
\usepackage[linesnumbered,ruled,vlined]{algorithm2e}%
\usepackage[table]{xcolor} 

\newtheorem{assumption}{Assumption}
\newtheorem{conjecture}{Conjecture}

\definecolor{lightgreen}{RGB}{200,255,200}
\definecolor{lightred}{RGB}{255,204,204}

\theoremstyle{thmstyleone}%
\newtheorem{theorem}{Theorem}

\theoremstyle{thmstyletwo}%

\theoremstyle{thmstylethree}%

\begin{document}

\title[Learning to Decode in Parallel]{Learning to Decode in Parallel: Self-Coordinating Neural Network for Real-Time Quantum Error Correction}

\author[1,3]{\fnm{Kai} \sur{Zhang}}
\equalcont{These authors contributed equally to this work.}

\author[4,5,6]{\fnm{Zhengzhong} \sur{Yi}}
\equalcont{These authors contributed equally to this work.}

\author[4,5,6]{\fnm{Shaojun} \sur{Guo}}

\author[2]{\fnm{Linghang} \sur{Kong}}

\author[1]{\fnm{Situ} \sur{Wang}}

\author[1]{\fnm{Xiaoyu} \sur{Zhan}}

\author[4,5,6]{\fnm{Tan} \sur{He}}

\author[4,5,6]{\fnm{Weiping} \sur{Lin}}

\author[4,5,6]{\fnm{Tao} \sur{Jiang}}

\author[4,5,6]{\fnm{Dongxin} \sur{Gao}}

\author[4,5,6]{\fnm{Yiming} \sur{Zhang}}

\author[3]{\fnm{Fangming} \sur{Liu}}

\author*[2]{\fnm{Fang} \sur{Zhang}}
\email{fangzhang@iqubit.org}

\author*[1,2]{\fnm{Zhengfeng} \sur{Ji}}
\email{jizhengfeng@tsinghua.edu.cn}

\author*[4,5,6]{\fnm{Fusheng} \sur{Chen}}
\email{fschen@hfnl.cn}

\author*[1]{\fnm{Jianxin} \sur{Chen}}
\email{chenjianxin@tsinghua.edu.cn}

\affil[1]{\orgdiv{Department of Computer Science and Technology}, \orgname{Tsinghua University}, \orgaddress{\city{Beijing}, \country{China}}}

\affil[2]{\orgname{Zhongguancun Laboratory}, \orgaddress{\city{Beijing}, \country{China}}}

\affil[3]{\orgname{Pengcheng Laboratory}, \orgaddress{\city{Shenzhen}, \state{Guangdong}, \country{China}}}

\affil[4]{\orgdiv{Hefei National Research Center for Physical Sciences at the Microscale and School of Physical Sciences}, \orgname{University of Science and Technology of China}, \orgaddress{\city{Hefei}, \country{China}}}

\affil[5]{\orgdiv{Shanghai Research Center for Quantum Science and CAS Center for Excellence in Quantum Information and Quantum Physics}, \orgname{University of Science and Technology of China}, \orgaddress{\city{Shanghai}, \country{China}}}

\affil[6]{\orgdiv{Hefei National Laboratory}, \orgname{University of Science and Technology of China}, \orgaddress{\city{Hefei}, \country{China}}}

\abstract{Fast, reliable decoders are pivotal components for enabling fault-tolerant quantum computation. Neural network decoders like AlphaQubit have demonstrated significant potential, achieving higher accuracy than traditional human-designed decoding algorithms. However, existing implementations of neural network decoders lack the parallelism required to decode the syndrome stream generated by a superconducting logical qubit in real time. Moreover, integrating AlphaQubit with sliding window-based parallel decoding schemes presents non-trivial challenges: AlphaQubit is trained solely to output a single bit corresponding to the global logical correction for an entire memory experiment, rather than local physical corrections that can be easily integrated. We address this issue by training a recurrent, transformer-based neural network specifically tailored for parallel window decoding. While our network still outputs a single bit, we derive training labels from a consistent set of local corrections and train on various types of decoding windows simultaneously. This approach enables the network to self-coordinate across neighboring windows, facilitating high-accuracy parallel decoding of arbitrarily long memory experiments.

As a result, we overcome the throughput bottleneck that previously precluded the use of AlphaQubit-type decoders in fault-tolerant quantum computation. Our work presents the first scalable, neural-network-based parallel decoding framework that simultaneously achieves state-of-the-art accuracy and the stringent throughput required for real-time quantum error correction. Using an end-to-end experimental workflow, we benchmark our decoder on the \textit{Zuchongzhi 3.2} superconducting quantum processor against existing human-designed decoders on surface codes with distances up to $7$, demonstrating its superior logical accuracy. Moreover, we demonstrate that, using our approach, a single TPU v6e is capable of decoding surface codes with distances as large as $25$ within $1$ $\mu s$ per decoding round.}


\maketitle

\section{Introduction}
\label{sec:introduction}

Inspired by Feynman's early vision, the theoretical advantages of quantum computing were firmly established three decades ago---evidenced by the quantum advantage of Shor's algorithm for factoring~\cite{shor1999polynomial}. Specifically, a system comprising several hundred perfect qubits could efficiently factor the $2048$-bit integers underlying RSA encryption~\cite{gidney2025factor}: a computational task that has long been believed to be intractable for classical computers.
However, in all physical implementations, whether based on superconducting~\cite{krantz2019quantum}, trapped-ion~\cite{bruzewicz2019trapped}, or other architectures~\cite{bluvstein2024logical, slussarenko2019photonic}, qubits are highly susceptible to decoherence caused by environmental perturbations.

To enable reliable quantum computation, quantum error correction codes (QECC) and fault-tolerant quantum computation schemes have been developed~\cite{gottesman1997stabilizer}. These techniques encode logical qubits into increasingly redundant ensembles of physical qubits, enabling higher logical fidelity as the code size grows---provided that physical error rates remain below the fault-tolerance threshold. Although it substantially increases hardware complexity by requiring two to three orders of magnitude more qubits, the fundamental value proposition remains: the benefits brought by the intrinsic speedup of quantum algorithms for problems like factoring can easily outweigh the overhead costs introduced by QECC.

\begin{figure}
    \centering
    \begin{subfigure}{0.3\linewidth}
        \includegraphics[width=\linewidth]{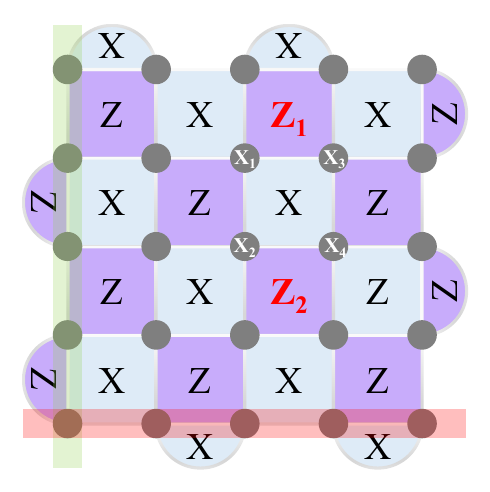}
        \caption{Surface code example.}
        \label{fig:surface_code}
    \end{subfigure}
    \begin{subfigure}{0.55\linewidth}
        \includegraphics[width=\linewidth]{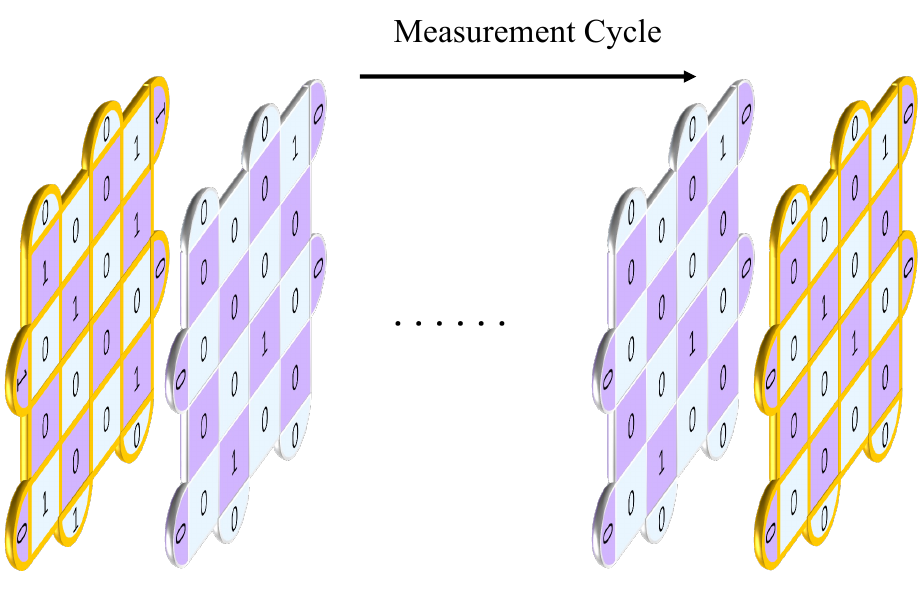}
        \caption{Memory experiment.}
        \label{fig:memory_experiment}
    \end{subfigure}
    \caption{(a) Surface code example for $d = 7$. Data qubits are represented by gray circles. Z stabilizers and X stabilizers are represented by the purple and blue squares (or semi-circle), respectively. Logical Z/X operators are illustrated as red/green rectangles. (b) Memory experiment example that contains multiple syndrome extraction rounds. We use light yellow shading to indicate the time boundaries of the initial and final rounds, as they are physically distinct from the intermediate rounds.}
\end{figure}

Among all coding schemes that rely on planar connectivity, the surface code~\cite{Dennis_2002} stands out as one of the most promising approaches, as it boasts the highest fault-tolerance threshold~\cite{aharonov1997fault, kitaev1997quantum, Knill1998ResilientQC} currently known. As illustrated in Fig.~\ref{fig:surface_code}, a surface code patch typically encodes one logical qubit into a $d \times d$ lattice of physical qubits. The code distance, denoted as $d$, is defined as the weight of the shortest logical operator (red line for the logical Z operator $Z_L$ and green line for the logical X operator $X_L$). As long as the physical error rate $p$ is below the fault-tolerance threshold, the logical error rate can be suppressed exponentially as the code distance increases, providing the desired accuracy for logical quantum computation.

The fault-tolerance threshold depends not only on the QECC design and the noise model, but also on the decoding algorithm. As a purely classical component, the decoder is tasked with outputting a result for each logical measurement in the circuit. To do so effectively, it must continuously monitor the \emph{error syndromes} generated at every step of the computation and track likely error configurations. \emph{Optimal decoding} corresponds to maximum-likelihood decoding (MLD)~\cite{Dennis_2002}, which, in theory, exhaustively evaluates all possible error configurations and selects the logical measurement result with the highest total probability. However, decoding the surface code near-optimally is a surprisingly difficult problem. Minimum weight perfect matching (MWPM)~\cite{Edmonds1973MatchingET, Kolmogorov2009BlossomVA} is a proposed heuristic solution that identifies a single most likely error configuration consisting only of X-type and Z-type errors. Although providing relatively high decoding efficiency, neglecting degenerate error configurations and X–Z correlated errors also prevents its accuracy from reaching the optimum.

The performance gap between MWPM and optimal decoding has been explicitly demonstrated for the repetition code, a low-dimensional analogue of the surface code that admits a polynomial-time maximum-likelihood decoder~\cite{cao2025exact}. For the surface code, optimal decoding remains computationally prohibitive except for the smallest code distances and number of rounds, yet existing high-accuracy decoders already confirm that MWPM is far from optimal. As early as 2013, Fowler~\cite{fowler2013optimal} proposed a simple modification to MWPM that incorporates correlation via reweighting. Although Fowler's correlated matching decoder yields only a modest improvement in threshold, it significantly enhances sub-threshold scaling—nearly doubling the benefit of increasing the distance from $d = 3$ to $d = 5$ at $p = 2 \times 10^{-4}$. Higgott~\cite{higgott2023improved} introduced belief-matching, which uses belief propagation for reweighting and improved the fault-tolerance threshold from $0.82\%$ to $0.94\%$ under their error model. Google Quantum AI~\cite{Acharya2022SuppressingQE} developed a tensor network decoder that significantly outperformed belief-matching on both experimental and simulated data, though at a cost of being many orders of magnitude slower. Shutty et al.~\cite{shutty2024efficient} designed Harmony, an ensembled correlated matching decoder that approaches the accuracy of tensor network decoders while being substantially faster and embarrassingly parallelizable.

When it comes to decoding real experimental data as opposed to simulated data generated from an error model, an additional challenge arises. Real quantum hardware contains many qubits and couplers with varying error rates, along with more subtle error sources such as crosstalk~\cite{tripathi2022suppression} and leakage~\cite{ghosh2013understanding}. These error processes are not fully understood~\cite{sivak2024optimization}, and they may not be perfectly captured by a simple hypergraph model that only contains the explicit noise sources in the quantum circuit. Although traditional decoding algorithms can often succeed with only an approximate error model despite these complications, this limitation still prevents them from fully harnessing the error-correcting potential of the code. 

The learning-based method has been an increasingly common approach to decoding quantum error correction codes. The pattern recognition capability of neural networks allow them to capture some hard-to-define features, and thus handle X-Z correlation and error configuration degeneracy better than traditional human-designed decoders. Furthermore, learning-based decoders can be fine-tuned on real hardware data, allowing them to take into account subtle characteristics of hardware errors that humans may not even know how to model. These two aspects together have contributed to a significant accuracy gap between state-of-the-art neural decoders and traditional decoders. Numerous efforts have been made to design neural decoders for surface codes~\cite{Ni2018NeuralND, liu2019neural, breuckmann2018scalable, Meinerz2021ScalableND, zhang2023scalable, hu2025efficient, cao2025generative}, with AlphaQubit~\cite{bausch2024learning} emerging as a leading example. It employs a recurrent transformer-based architecture and is trained to predict the global logical correction in memory experiments, allowing for fine-tuning with real hardware data. AlphaQubit achieves state-of-the-art performance on Sycamore experimental data for $d = 3$ and $d = 5$, and outperforms correlated matching on simulated data up to $d = 11$.

However, AlphaQubit also has some bottlenecks that need to be solved before it becomes plausible to use for large-scale fault-tolerant quantum computation. One important bottleneck is its speed---more precisely, the decoding \emph{throughput}---which affects many traditional decoder too, but is especially problematic for AlphaQubit. All known decoders that exceed MWPM in accuracy incur higher computational cost, reflecting a general trade-off between decoding speed and performance. This poses a critical challenge for fault-tolerant quantum computation, which requires real-time decoding~\cite{Zhang2023ACA}. Many decoders mitigate this cost through parallelization~\cite{Tan2022ScalableSD, Skoric2022ParallelWD, bombin2023modular}, increasing the total computational overhead by only a constant factor while enabling distribution across an arbitrary number of decoding units. However, such schemes depend on merging results from adjacent decoding windows using local physical error predictions. By contrast, AlphaQubit produces only global logical predictions and therefore does not admit this form of parallelization.

In this work, we seek to train a neural decoder that retains the accuracy of AlphaQubit while supporting parallel decoding. We observe that, when applying parallel decoding to MWPM, the merge step is often not actually executed at all, since outputs from adjacent windows are already consistent with each other at the seam. Indeed, as the size of the overlap region between windows increases, the probability that the merge step is required diminishes exponentially, eventually becoming negligible compared to the logical error rate of the code. 

Motivated by this observation, we train a neural network to output a logical correction bit for each window that can be combined directly, without local merging. Although there is no unique ``correct'' decoding of the core region of an individual window, we can train the neural network so that outputs from adjacent windows are usually ``consistent'' with respect to how shared syndromes are handled, even though we can no longer define ``consistent'' exactly like we could with traditional matching-based decoders. Such consistency is sufficient to ensure that, when the per-window outputs are combined into a logical outcome, the accuracy of the base decoder is preserved.

\section{Background}
\label{sec:background}
\subsection{Real-time decoding}
\label{sec:real_time_decoding}
A significant fraction of errors in quantum computers arise from decoherence, which occurs regardless of whether the qubits are undergoing gates or idling. Consequently, when implementing fault-tolerant quantum computation, the quantum computer cannot just pause and wait for the decoding result: without periodic syndrome extraction rounds to help locating and isolating decoherence errors, they would accumulate on the data qubits and quickly go over the threshold of the fault-tolerant protocol. In particular, a superconducting quantum computer can execute a round of syndrome extraction every $\sim 1 \mu s$~\cite{RyanAnderson2021RealizationOR}, and considering that the currently demonstrated noise level is barely below threshold~\cite{google2025quantum}, we really do not want to go any slower.

Fortunately, the quantum computer rarely \emph{needs} to wait for the decoding result. Thanks to the Pauli frame technique~\cite{Riesebos2017PauliFF, chamberland2018fault, knill2005quantum}, there is no need to correct every physical error individually and immediately. Instead, the corrections can be tracked by the decoding unit, and applied only to the results of logical measurements~\cite{Zhang2023ACA}. Therefore, the decoder can only stall the quantum computer when the result of an earlier logical measurement is needed to determine what operation needs to be applied now, which happens, for example, when implementing non-Clifford gates with gate teleportation~\cite{litinski2019game}. This problem can be further mitigated by trying to schedule logical operations so that the quantum computer has other tasks to perform while waiting for the result. In the worst case, the quantum computer can stall \emph{on the logical level}, not performing any logical operations but continuing to run the syndrome extraction.

\begin{figure}
  \centering
  \includegraphics[width=0.5\linewidth]{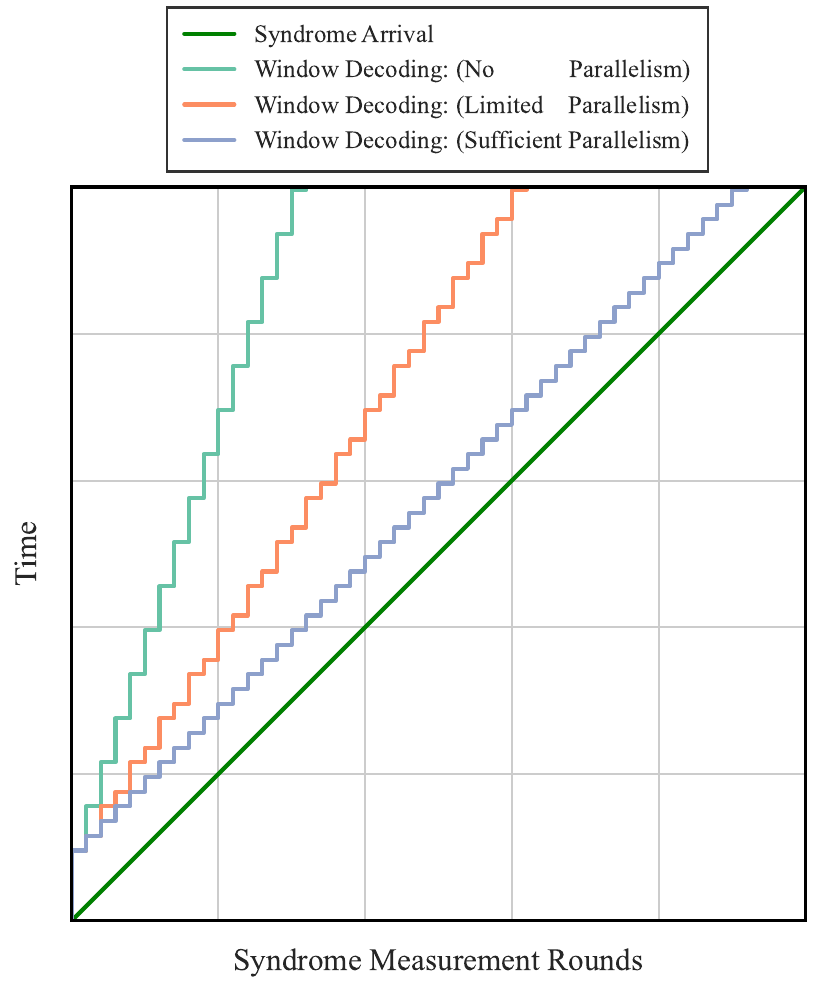}
  \caption{Influence of decoding parallelism. The decoding latency is the time difference between the decoder response and the syndrome generation. For parallel decoders with insufficient throughput, the latency also grows nearly linearly and becomes unacceptable. In contrast, for a decoder with sufficient throughput, the latency remains asymptotically constant regardless of the number of measurement rounds.}
  \label{fig:realtime_decoding}
\end{figure}

Nonetheless, there is a minimum requirement on the speed of the decoder. If the average speed at which the decoder can process syndromes, the decoding \emph{throughput}, is slower than the rate at which the quantum computer generates them, the decoder will fall progressively further behind, as illustrated in Fig.~\ref{fig:realtime_decoding}. The problem is exacerbated by the fact that, as mentioned earlier, the quantum computer continues to generate syndromes even while stalled. Moreover, syndromes generated during a stall for one logical measurement may need to be decoded to determine the \emph{next} logical measurement. If the stall duration increases linearly with the total number of rounds so far, the number of rounds per logical measurement in a dependency chain can grow exponentially~\cite{Terhal2013QuantumEC}, potentially negating any quantum advantage over classical computation. Unfortunately, AlphaQubit~\cite{bausch2024learning} operates in this regime. Although its recurrent architecture allows it to process syndromes as soon as they become available, each round still takes $20 \ \mu s$ even at $d = 3$, too slow for a superconducting quantum computer operating at $1 \ \mu s$ per round. 

One strategy to avoid exponential stalling with AlphaQubit is solely through low-level optimizations such as quantization~\cite{jacob2018quantization} and pruning~\cite{han2015learning}, which may be able to reduce the single-round inference latency to below 1 $\mu s$. Under such a strategy, the inference speed of the classical hardware remains a hard limiting factor, and increasing the code distance $d$ or the complexity of the neural network may easily bring the inference latency to over 1 $\mu s$ again. In contrast, if a \emph{parallelized} AlphaQubit implementation were available, where inference can be executed in parallel, then any throughput requirement can be met simply by increasing the parallelism. Even if the streaming inference latency remains over 1 $\mu s$, it could be maintained at a constant level, as already demonstrated by Willow~\cite{google2025quantum}. This constant can be further reduced by implementing the aforementioned low-level optimizations, thus improving the logical feedback speed and the fidelity in lattice surgery~\cite{litinski2019game}.

In other words, a decoder with sufficient throughput does not necessarily eliminate stalling, but keeps it under control. There will still be a \emph{latency} between the last input syndromes and the final decoding results, but over a long computation session, this latency remains asymptotically constant as illustrated in Fig.~\ref{fig:realtime_decoding}. This means that for every logical measurement, the decoder will at most stall the computation for a constant amount of time, and the overall time cost will remain linear in the size of the quantum circuit.
A prime example is the parallel sliding-window decoder~\cite{Tan2022ScalableSD, Skoric2022ParallelWD, bombin2023modular}. This decoder necessarily incurs a large initial latency, both to collect enough syndromes to fill a window and to decode them. While parallelization between windows cannot reduce this initial latency, sufficient parallelism ensures that all subsequent windows are processed with effectively the same latency.

\begin{figure}
  \centering
  \includegraphics[width=\linewidth]{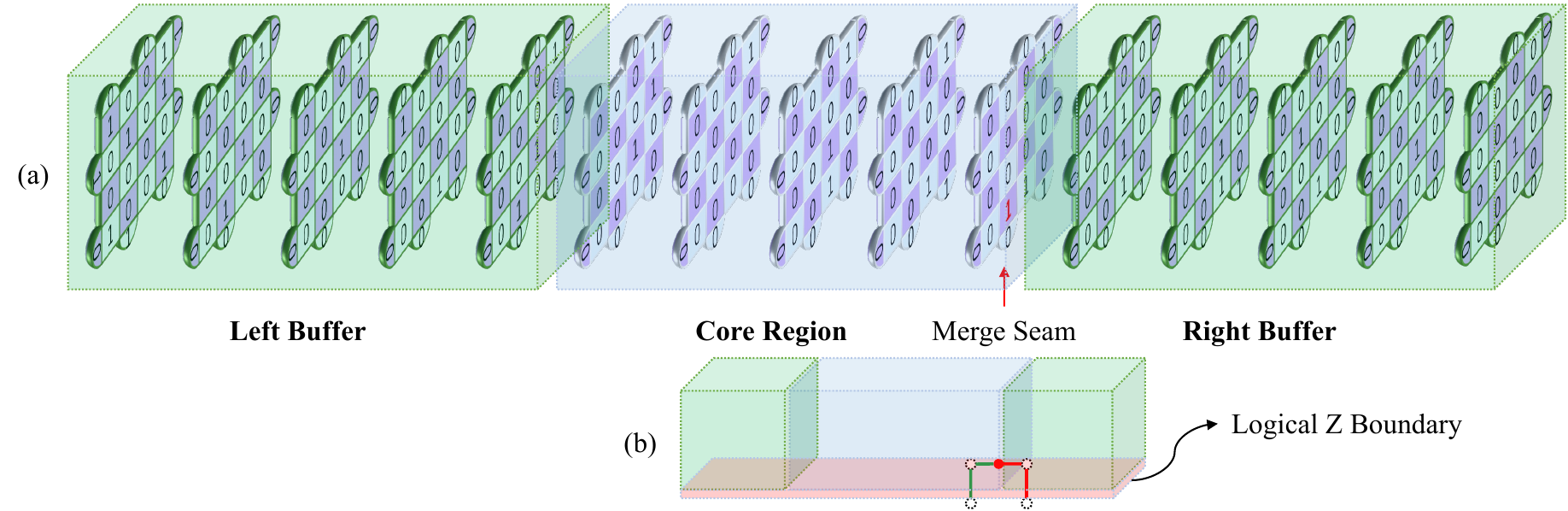}
  \caption{Decoding window visualization for $d = 5$ and degeneracy example in merge seam with logical error in the core region or the buffer region.}\label{fig:decoding_window}
\end{figure}

\subsection{Parallel decoding without local merging}
\label{sec:without_merging}

Existing implementations of the parallel sliding-window decoder~\cite{Tan2022ScalableSD, Skoric2022ParallelWD, bombin2023modular} are usually framed in a hypergraph-based model of decoding. In such a model, a decoding task is represented by a hypergraph $G = (V, E)$, where each edge $e \in E$ corresponds to a physical error source, and each vertex $v \in V$ corresponds to a \emph{detector}, a bit derived from measurement results that indicates the presence of nearby errors. The value of a detector is the parity of the number of errors associated with it that actually occurred. In other words, each error flips all associated detectors. 

A notable characteristic of quantum codes is that \emph{degeneracy} is common. There will be many short \emph{cycles}, i.e., small sets of edges $C \subseteq E$ that flip each vertex an even number of times, which represent low-weight \emph{undetectable errors}. However, these undetectable errors do not undermine the fault tolerance of the surface code because they also act trivially on the encoded logical qubit. Concretely, we can define a set of edges $L \subseteq E$ that represents a logical operator, such that a set of errors $\mathcal{E} \subseteq E$ flips this logical operator if and only if $|\mathcal{E} \cap L| \equiv 1 \pmod{2}$. An undetectable error that flips a logical operator must have weight at least $d$, enabling the exponential suppression of logical error rates.

One consequence of the ubiquity of degeneracy is that there often exist many equivalent ways to correct the same error syndrome, as shown by Fig.~\ref{fig:surface_code}: both error sets $\{X_1, X_2\}$ and $\{X_3, X_4\}$ will flip the same stabilizers $Z_1$ and  $Z_2$, but that is fine because both corrections have the same effect on the logical operator.

For sliding-window decoders, the basic idea is to break the decoding graph $G$ into different sub-graphs, or decoding windows~\cite{Tan2022ScalableSD}, such that each decoding process only makes the decision on a subset of edges, called the \emph{core region} of a window. To ensure the accuracy, each window contains not only the detector information within the core region, but also a \emph{buffer region} around the core region, as shown in Fig.~\ref{fig:decoding_window}. Degeneracy can become a problem here because adjacent windows may output two sets of corrections that are globally equivalent, but not equivalent when the boundary of the core region is considered. Suppose there is a Z detector flip (illustrated as the red vertex in Fig.~\ref{fig:decoding_window}\textcolor{blue}{a}) on the right boundary of the core region, i.e., the right merge seam for this decoding window. Then the 
ground truth correction can be an error chain that connects this detector to the logical Z boundary in the core region (the green error chain), but also an error chain that connects to the logical Z boundary in the right buffer (the red error chain). This will result in a logical error if both the current and the next decoders output a logical correction in the core region (or both ignore the correction), since the ground truth logical operator corresponding to this error configuration should be flipped only once.

Ref.~\cite{Tan2022ScalableSD} handles this problem by \emph{merging}
the corrections at the \emph{seam} between core regions, combining the local corrections within each window's core region and decoding them again on a 2D decoding graph.
Ref.~\cite{Skoric2022ParallelWD} also employs two rounds of decoding; although
the seam decoding graphs are extended into 3D windows with fixed boundary
conditions, the underlying principle remains the same.
Ref.~\cite{bombin2023modular} further shortens the core regions in the first
round of windows, so that the primary purpose of the initial decoding round
becomes determining an appropriate set of boundary conditions for the second
round.

Despite these differences, all these schemes require the base decoder to output a set of local corrections in order to enforce consistency between the outputs of adjacent windows. This is incompatible with AlphaQubit, which does not rely on explicit knowledge of the decoding graph and outputs only a single bit for global logical correction.

Our novel observation is that, when using the scheme of~\cite{Tan2022ScalableSD}, as long as the length of the buffer region is sufficiently large, the probability of any remaining flipped detectors requiring merging becomes extremely low compared to the logical error rate. As we will show in Appendix~\ref{sec:appendix_decoding_without_merging}, for the memory experiment with code distance $d$, the minimum length of the buffer region required to preserve the fault-tolerant distance without merging is $d$.

This observation inspires us to train a neural network decoder whose output can be used in the parallel decoding scheme without merging. Similar to AlphaQubit, our decoder outputs only a single bit of logical correction per window. As long as these outputs are consistent across windows, they can be simply XORed together to obtain a global correction for the entire experiment.

In the following sections, we will describe our model, including detailed explanations of its architecture, a novel supervision framework tailored to parallel window decoding, and the corresponding training methodology.

\begin{figure}
\centering
\includegraphics[width=\linewidth]{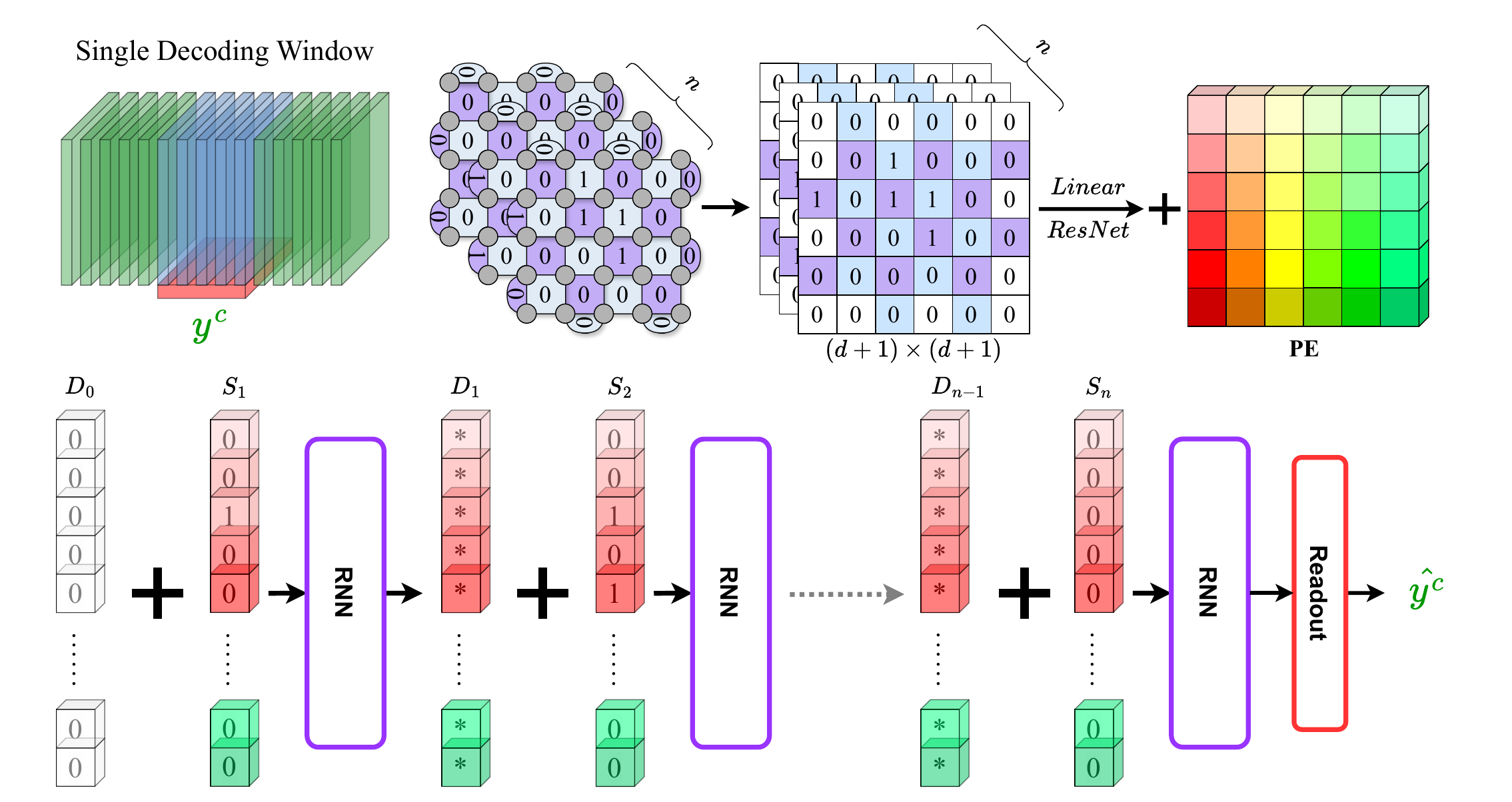}
\caption{Overview of our parallel decoding scheme. Each decoding window containing $n$ syndrome measurement rounds will be embedded and go through the RNN for $n$ times, exactly the same as AlphaQubit, while the final output logit is designed to predict the logical error only in the core region of this decoding window.}
\label{fig:overview}
\end{figure}

\section{Methods}
\label{sec:method}
\subsection{Parallel neural decoding scheme}
\label{sec:parallel_neural_decoding}
In the parallel decoding scheme, the input syndrome stream from the quantum hardware is divided into overlapping sliding windows. Our neural network model takes syndromes from one decoding window as input, and outputs a single bit representing the contribution of the core region of this window to the overall logical correction, as illustrated in Fig.~\ref{fig:overview}. Since the first and last  decoding windows are slightly different in the memory experiment due to the temporal boundary, we pad them appropriately following the strategy described in Appendix~\ref{sec:Appendix_B}, so that they can be handled by the same network and trained simultaneously.

To enable the model to predict logical errors only in the core region of a decoding window, we define the \emph{local ground truth} $\mathcal{E}$ as the set of edges in the decoding graph which have been flipped, representing the underlying physical errors. This cannot be observed in real experiments, but is available in simulation. The ground truth for each decoding window during the training process is then derived as:
\begin{equation}\label{eq:core_logical_error}
  y_i = |\mathcal{E} \cap E^c_i \cap L| \bmod 2
\end{equation}
where $E^c_i$ is the set of edges in the core region of window $i$, and $L$ is a global logical operator. Since $\{E^c_i\}$ is a partition of the original $E$, it is guaranteed that:
\begin{equation}
    \bigoplus_{i=1}^{m} y_i = y = |\mathcal{E} \cap L| \bmod 2
    \label{eq:global_xor}
\end{equation}
where $m$ is the number of decoding windows split from the global decoding graph. In other words, if the neural network correctly predicts $y_i$ for each window $i$, then combining all predictions with XOR yields the correct overall logical correction. 

Note that the converse is not necessarily true: if the network mis-predicts an even number of $y_i$, the overall prediction $y$ would still be correct. This means that, if there are two degenerate error configurations (as depicted in Fig.~\ref{fig:decoding_window}\textcolor{blue}{b}), the network can \emph{consistently} predict any one of them to arrive at the same correct overall prediction. However, inconsistent predictions between adjacent windows will cause a logical error. As demonstrated later by subsequent experiments, our strategy of joint training all decoding windows help the decoder to output consistent results in most cases.

\begin{figure}
\centering
\includegraphics[width=\linewidth]{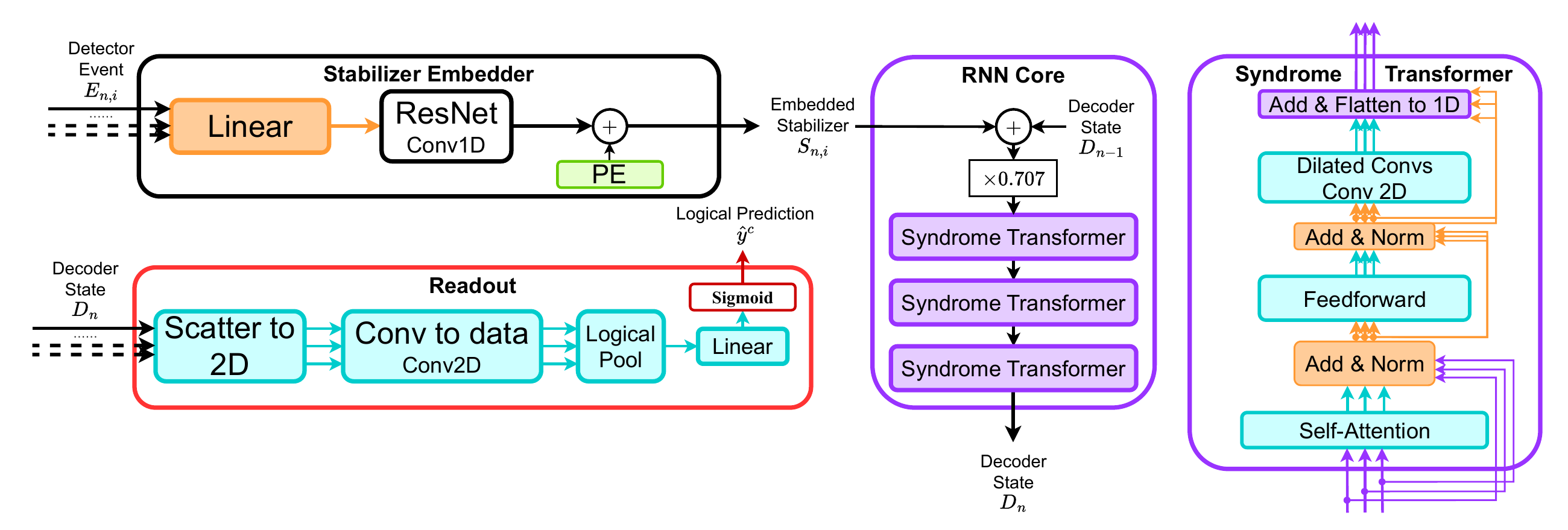}
\caption{Neural network architecture. The basic network architecture is consistent with that of AlphaQubit, with the key modification being that the supervisions are changed from the global logical error to the local logical error within a decoding window.}
\label{fig:neural_network_architecture}
\end{figure}

\subsection{Model design and implementation}
The architecture of our neural network is depicted in Fig.~\ref{fig:neural_network_architecture}. We adopt the model design of AlphaQubit, but our model only takes hard detector events as input for simplicity. However, adding syndrome measurement, soft readout and leakage information~\cite{Acharya2022SuppressingQE, varbanov2025neural, bausch2024learning} is possible and may improve performance in the future. There are also several simplifications for practical implementation. The subtle differences from the original model design in AlphaQubit and the corresponding considerations are detailed in Appendix~\ref{sec:Appendix_B}, and will be discussed in our ablation study.

\subsection{Pretraining in simulation}
\label{sec:pretraining}
To train the model from scratch, we train on simulated noise data generated by Stim~\cite{Gidney2021StimAF}. We build up the surface code circuit with the same circuit-level noise model as in~\cite{Tan2022ScalableSD}: each reset operation intending to set a qubit to $\ket{0}$ (resp. $\ket{+}$) has probability $p$ to set it to $\ket{1}$ (resp. $\ket{-}$) instead, each measurement has probability $p$ to get the result flipped, and each two-qubit gate and idle gate is followed by a depolarizing channel with strength $p$. Idle gates are inserted on qubits near the boundary not participating in a two-qubit gate, and on all data qubits twice while ancilla qubits are undergoing measurement and then reset.

\subsubsection{Singular prediction training}
As the logical error prediction is exactly a binary classification problem, we use the binary cross entropy (BCE) loss for all kinds of decoding windows:

\begin{equation}
Loss = - \frac{1}{B} \sum_{i=1}^{B} \left[y^c_{i} \log(\hat{y}^c_{i}) + (1 - y^c_{i}) \log(1 - \hat{y}^c_{i}) \right]
\label{eq:loss_func}
\end{equation}
where $B$ is the batch size, $y^c_i$ is the ground truth logical error flip in the core region of a decoding window, acquired from the simulator~\cite{Gidney2021StimAF}, and $\hat{y}^c_{i}$ is the window prediction.

\begin{algorithm}[th]
\caption{Parallel Neural Decoding}
\label{alg:rnn_decoder_alg}
\KwIn{$\mathbf{E} \in \mathbb{R}^{B \times (b + c + b) \times (d + 1) \times (d + 1)}$: Input detector events}
\KwOut{$\hat{y}$: Logical predictions without or with recurrent training strategy}
\BlankLine
Initialize $decoder\_state \ \mathbf{D_0} \gets \mathbf{0} \in \mathbb{R}^{B \times (d + 1)^2 \times d_{model}}$\;
\If{not recurrent\_training}{
    \For{$n \gets 1$ \KwTo $b + c + b$}{
        $\mathbf{E_n} \gets \mathbf{E}[:, n]$\;
        $\mathbf{S_n} \gets \textbf{StabilizerEmbedder}(\mathbf{E_n})$\;
        $\mathbf{D_n} \gets \textbf{RNNCore}(\mathbf{S_n}, \mathbf{D_{n-1}})$\;
    }
    $\hat{y} \gets \textbf{Readout}(\mathbf{D_n})$\;
    \Return{$\hat{y}$}\;
}
\Else{
    $\hat{y} \gets [\ ]$\;
    \For{$n \gets 1$ \KwTo $b + c + b$}{
        $\mathbf{E_n} \gets \mathbf{E}[:, n]$\;
        $\mathbf{S_n} \gets \textbf{StabilizerEmbedder}(\mathbf{E_n})$\;
        $\mathbf{D_n} \gets \textbf{RNNCore}(\mathbf{S_n}, \mathbf{D_{n-1}})$\;
        \If{$n > 2 * b$}{
            $\hat{y}_{single} \gets \textbf{Readout}(\mathbf{D_n})$\;
            Append $\hat{y}_{single}$ to $\hat{y}$\;
        }
    }
    \Return{$\hat{y}$}\;
}
\end{algorithm}

\subsubsection{Multi-layer recurrent training}
\label{sec:recurrent_training}
To facilitate the training process and guide the network to progressively learn to decode the core region, we design a recurrent training method in Algorithm~\ref{alg:rnn_decoder_alg}, which essentially allows the network to simultaneously train with different truncated core sizes $\tau = 1, 2, \dots, c$. This is especially useful for training large code distances, as learning to decode $b+c+b$ rounds of syndrome directly may be too challenging.

In the recurrent training mode, for each $\tau = 1, 2, \dots, c$, the first $b+\tau+b$ rounds of data in a decoding window are treated as if they were a complete decoding window with $b$ rounds of left buffer, $\tau$ rounds of core region, and $b$ rounds of right buffer. Therefore the middle $\tau$ rounds of ground truth are used to derive a label $y_i^\tau$, and the neural network also predicts a $\hat{y}_i^\tau$. Thanks to the recurrent structure of the RNN core, $\hat{y}_i^\tau$ can be computed by simply truncating the intermediate decoder state $D_{b+\tau+b}$ to the readout module, and thus all $\hat{y}_i^\tau$ can be computed with the same $b+c+b$ invocations of the RNN core. The loss formula of a batch becomes:

\begin{equation}
Loss = - \frac{1}{Bc} \sum_{i=1}^{B} \sum_{\tau=1}^{c} \left[y^\tau_i \log(\hat{y}^\tau_i) + (1 - y^\tau_i) \log(1 - \hat{y}^\tau_i) \right].
\end{equation}

\subsection{Hardware-aware fine-tuning}
\label{sec:fine-tuning}
After pretraining on simulated data, we further fine-tune the model in a two-stage procedure under realistic hardware decoding scenarios---decoding for the surface code memory experiment on the superconducting quantum processor.

Experimental attempts indicate that pretraining is almost indispensable, as it paves the way for real-device fine-tuning. If training the model directly on the hardware-aware detector error model (DEM), the complexity of realistic noise hinders the network's optimization, making convergence challenging. However, by first performing large-scale pretraining on the hardware-agnostic simulated data, which is generated from a much simpler noise model, the subsequent hardware-aware fine-tuning only needs to adapt to the shift in the noise distribution, rather than relearning the entire decoding process. Similar ideas have also been explored in modern large language models~\cite{brown2020language, ouyang2022training, hu2022lora}.

\subsubsection{DEM-based fine-tuning}
Before performing end-to-end fine-tuning on limited hardware data, we first calibrate our pretrained decoder using the DEM extracted from experimental data, as described in Appendix~\ref{sec:Appendix_C}. Compared to our simulated noise in the pretraining stage, the DEM is calculated using the correlation analysis method~\cite{chen2022calibrated, google2021exponential, Acharya2022SuppressingQE} directly by the hardware data, capturing more device-specific error characteristics.

The training loss function remains to be Equation~\ref{eq:loss_func}, but is performed on much fewer training samples with a reduced learning rate $\sim 10^{-5}$. This prevents overfitting while allowing the model to quickly adapt to the statistics of realistic noise processes. The DEM fine-tuning  provides a clear performance improvement over the model trained solely on hardware-agnostic simulation data in Section~\ref{sec:pretraining}.

\subsubsection{End-to-end hardware fine-tuning}
As mentioned in Section~\ref{sec:introduction} before, the power of neural network decoders, especially for AlphaQubit, comes from the end-to-end capability to infer hardware noise directly from the experimental data. Real quantum memory experiments differ from simulated ones in that the local ground truth $\mathcal{E}$ is not available, and only at the end of each run of the experiment we get the global ground truth $y = |\mathcal{E} \cap L| \bmod 2$. For the original AlphaQubit, this is not a problem because $y$ is the only label required for training, but our training method in Section~\ref{sec:pretraining} is not fully end-to-end. It needs the ground truth for each individual window $y_i$, and thus cannot be applied to real experimental data.

The loss function defined in Equation~\ref{eq:loss_func} relies on the window prediction $\hat{y}^c$ being a continuous value in the range $(0, 1)$, representing the confidence of the decoder about its prediction. To conduct end-to-end fine-tuning using experimental data directly, we need to convert the decoder predictions $\hat{y_i} \in (0, 1)$ for each individual window to a global prediction $\hat{y} \in (0, 1)$ in a differentiable way. To this end, we design a differentiable Soft-XOR module similar to~\cite{ozolins2022goal} to characterize the global XOR process under the group isomorphism $\{\{0, 1\}, \oplus\} \to \{+1, -1\}, \times\}$:
\begin{equation}
    \hat{y} \;=\; \tfrac{1}{2}\Bigl(1 - \prod_{k=1}^m \bigl(1 - 2\hat{y}^c_k \bigr)\Bigr)
    \label{eq:soft_xor}
\end{equation}
where $\hat{y}^c_k$ is the predicted probability of logical error in the $k$-th decoding window and $m$ is the number of decoding windows. 
This formulation is equivalent to a soft relaxation of the parity check across windows and enables end-to-end optimization through standard backpropagation, without changing any model architecture. The final loss function turns is obtained by changing the $\hat{y}^c$ and $y^c$ in Equation~\ref{eq:loss_func} to $\hat{y}$ and $y$, corresponding to the global prediction and global logical state, respectively.

\section{Results}
\label{sec:results}
In this section, we evaluate the proposed parallel decoding scheme across several key dimensions: \textbf{decoding accuracy}, \textbf{throughput}, and \textbf{scalability}. 

First, we report the decoding accuracy of the proposed scheme on both simulated data and experimental data from superconducting qubits in Section~\ref{sec:decode_accuracy}, demonstrating its clear advantage over human-designed decoders.

As discussed earlier, for a decoder to be viable in fault-tolerant quantum computing, its throughput must satisfy stringent requirements---namely, completing decoding within $1 \mu s$ per syndrome round under current assumptions for superconducting qubit systems. To assess this, we benchmark our implementation on a TPU v6e (not the most advanced hardware available). Remarkably, a single TPU v6e chip can support real-time decoding of the surface code up to distance $25$, even without further engineering optimizations. This significantly surpasses the real-time decoding capabilities of prior neural decoders: even AlphaQubit 2~\cite{senior2025scalable}, which was introduced just a few months after our theoretical proposal~\cite{zhang2025learning}, is limited to surface code distances of approximately $d=11$ on the same hardware (AQ2-RT).

We report the training costs for surface code distances $d=3$, $5$, and $7$, which were used in our experimental demonstrations. Extensive training for large-distance codes was beyond the scope of our current resources in an academic setting; nevertheless, we anticipate that the model architecture exhibits no intrinsic barriers to scaling.

In the following subsections, we discuss each of these aspects in turn.

\subsection{Decoding accuracy}\label{sec:decode_accuracy}
We evaluate the decoding accuracy of our neural network decoder against leading traditional decoding algorithms, including PyMatching~\cite{higgott2025sparse, Higgott2021PyMatchingAP}, Correlated-Matching~\cite{fowler2013optimal} and Belief-Matching~\cite{higgott2023improved}, under both simulated and hardware experiments. PyMatching represents the state-of-the-art pure MWPM decoder, with an exact implementation of MWPM and almost $\mu s$ level decoding latency on modern CPUs, making it a strong candidate for real-time quantum error correction. Correlated-Matching improves the decoding accuracy of MWPM by reweighting the X/Z decoding graphs to account for correlated errors, and has been integrated into the open-source library of Pymatching recently~\cite{higgott2025sparse}. Belief-Matching is a combination of belief-propagation (BP) and MWPM, with higher accuracy than pure MWPM but much lower decoding speed. For all traditional decoders, we keep their default configurations and decode the syndrome on the global decoding graph. All of them utilize the same DEMs that the neural network was trained on as the decoding prior to ensure fairness. For all our experiments, we set both the buffer region size $b$ and the core region size $c$ to the code distance $d$ to keep the parallelization overhead constant.

\begin{figure}[htbp]
\begin{center}
\includegraphics[width=\linewidth]{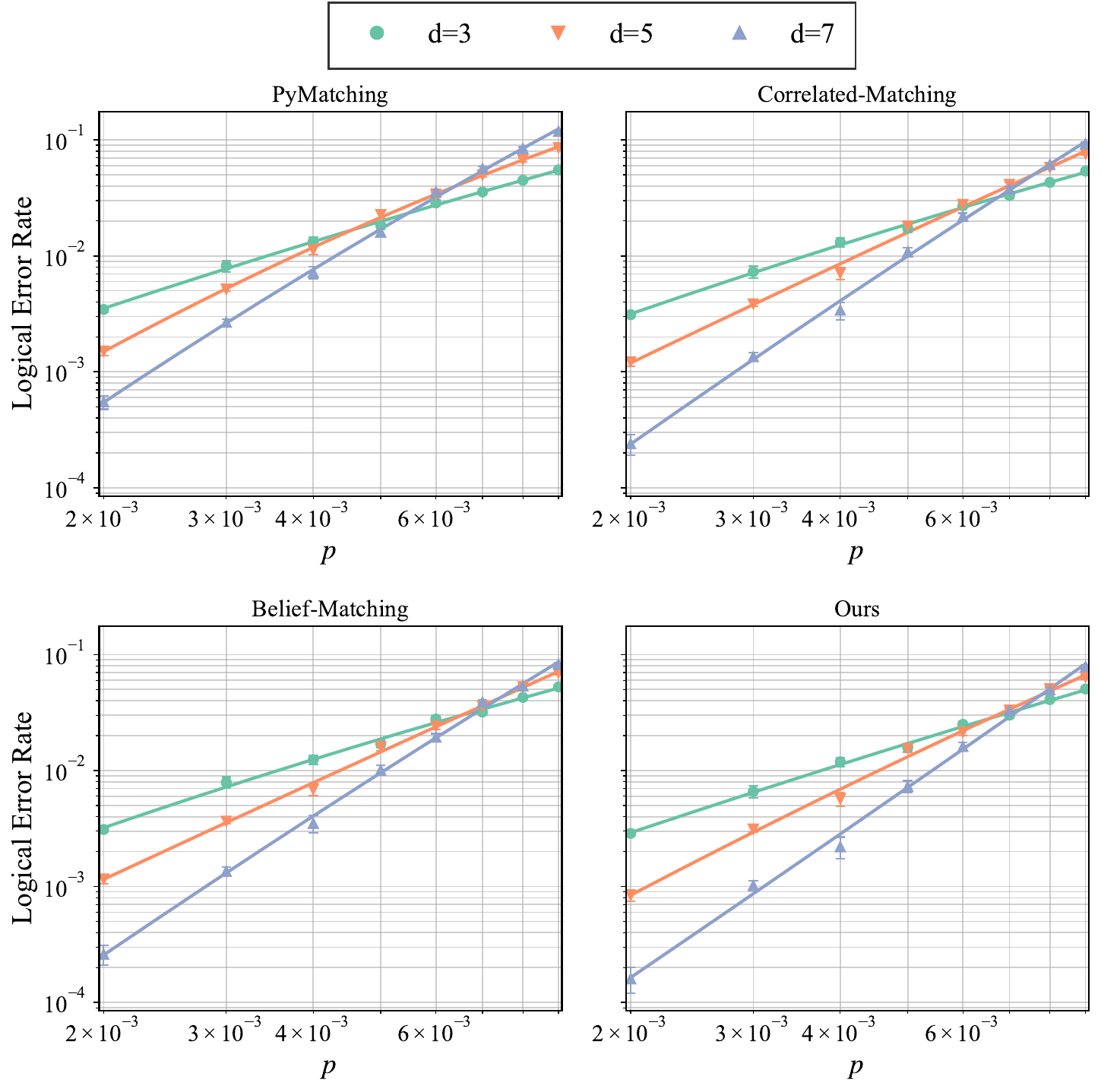}
\caption{Logical error rate of the standard memory experiment with $d$ rounds of syndrome measurement under the circuit-level noise model. The physical error rate $p$ is varied from $2 \times 10^{-3}$ to $9 \times 10^{-3}$. The fault-tolerance threshold is estimated from the intersection points of the logical error rate curves at different code distances, yielding threshold values of approximately $0.62\%$ for PyMatching, $0.73\%$ for Belief-Matching, $0.74\%$ for Correlated-Matching, and $0.76\%$ for our neural decoder.}
\label{fig:threshold_all}
\end{center}
\end{figure}

\subsubsection{Decoding under simulated noise}

We first conduct the standard memory experiment with $d$ syndrome measurement rounds to demonstrate our neural decoder's accuracy and the improvement of the fault-tolerance threshold. Since our model is designed for parallel window decoding and is tested for at least one decoding window, we pad the original syndrome with a left buffer of $d$ zero-valued rounds and a right buffer of $d$ zero-valued rounds, forming a single window. As shown in Fig.~\ref{fig:threshold_all}, compared with PyMatching, Correlated-Matching and Belief-Matching, our decoder achieves the lowest logical error rates at each physical error rate $p$ and the highest fault-tolerance threshold (approximately $\mathbf{0.76\%}$ under our noise model).
This indicates that the accuracy of our decoder is comparable with state-of-the-art neural decoders like AlphaQubit, even though our decoder is trained with per-window labels that are more noisy due to the degeneracy problem, and has to learn to handle the core region and the buffer region differently.

To study how well the neural network predicts $y_i$ for each window $i$, we simulate memory experiments for $d = 3, 5, 7$ with $N = 3d$ measurement rounds, which will be divided into exactly one start, one bulk, and one final decoding window by our decoding scheme. We feed all decoding windows into our neural network to get predictions $\hat{y}_i$, and compare them with $y_i$ to get a ``logical error rate'' $p_i = \Pr[\hat{y}_i \ne y_i]$ for each type of windows. We also calculate the ground-truth global logical error rate $p_g = \Pr[\hat{y}_1 \oplus \hat{y}_2 \oplus \hat{y}_3 \ne y_1 \oplus y_2 \oplus y_3]$. The results are presented in Table~\ref{tab:independent_decoding}.

\begin{table}[h]
\centering
\caption{Logical error rates for the individual windows and the entire $N = 3d$ memory experiment and $n=3$ windows. The column labeled $\hat{p}_g$ is the estimated global logical error rate from the error rates of each window, assuming that those ``errors'' happen independently between windows.}
\label{tab:independent_decoding}
\begin{tabular}{c  c  ccccc}
\toprule
\multirow{3}{*}{$d$} & \multirow{3}{*}{$p$} & \multicolumn{5}{c}{Logical Error Rate} \\
\cline{3-7}
 &  & $p_1$ & $p_2$ & $p_3$ & $p_g$ & $\hat{p}_g$\\
 &  & Initial Window & Bulk Window & Final Window & Ground-Truth & Estimated\\
\midrule
3 & 0.003 & 0.0140 & 0.0188 & 0.0120 & \textbf{0.0294 $\pm$ 0.0022} & 0.0435\\
5 & 0.003 & 0.0112 & 0.0172 & 0.0124 & \textbf{0.0130 $\pm$ 0.0017} & 0.0398\\
7 & 0.003 & 0.0116 & 0.0212 & 0.0114 & \textbf{0.0038 $\pm$ 0.0005} & 0.0430\\
\bottomrule
\end{tabular}
\end{table}
We observe that these results show that these ``mispredictions'' are not independent between windows. If they were, then the global logical error rate should be given by:
\begin{equation}
    \hat{p}_g = p_1 (1-p_2)(1-p_3) + (1-p_1) p_2 (1-p_3) + (1-p_1)(1-p_2) p_3 + p_1 p_2 p_3.
    \label{eq:independent_decoding}
\end{equation}

However, in Table~\ref{tab:independent_decoding}, we see that $p_g$  is consistently lower than $\hat{p}_g$. Furthermore, as $d$ increases, $p_g$ is suppressed rapidly while $\hat{p}_g$ does not change much. When $d=7$, most individual window mispredictions do not cause global logical errors, but instead are paired up with each other and eliminated in the XOR operation.

We postulate that these mispredictions are caused by degeneracy at the seam between windows as discussed in Section~\ref{sec:without_merging}. When two decoding windows handle a syndrome configuration in a way that is different from the ground truth, but topologically equivalent to the ground truth and \emph{consistent} with each other, both windows would mispredict their own label $y_i$, but the combined global prediction would remain correct. This happens more often for the bulk window since it has two seams instead of one, as confirmed by our data.

This also indicates that the neural network does not learn the errors of individual windows completely independently but rather develops a certain self-coordination among them.
The self-coordination enables our decoder to employ a parallel decoding scheme that generalizes to an arbitrary number of syndrome measurement rounds: each decoding thread decodes a single decoding window in a streaming manner, and the 1-bit logical results produced by different threads are ultimately gathered and XORed to obtain the final logical correction.

\begin{figure}[!htbp]
\begin{center}
    \begin{subfigure}{\linewidth}
        \includegraphics[width=\linewidth]{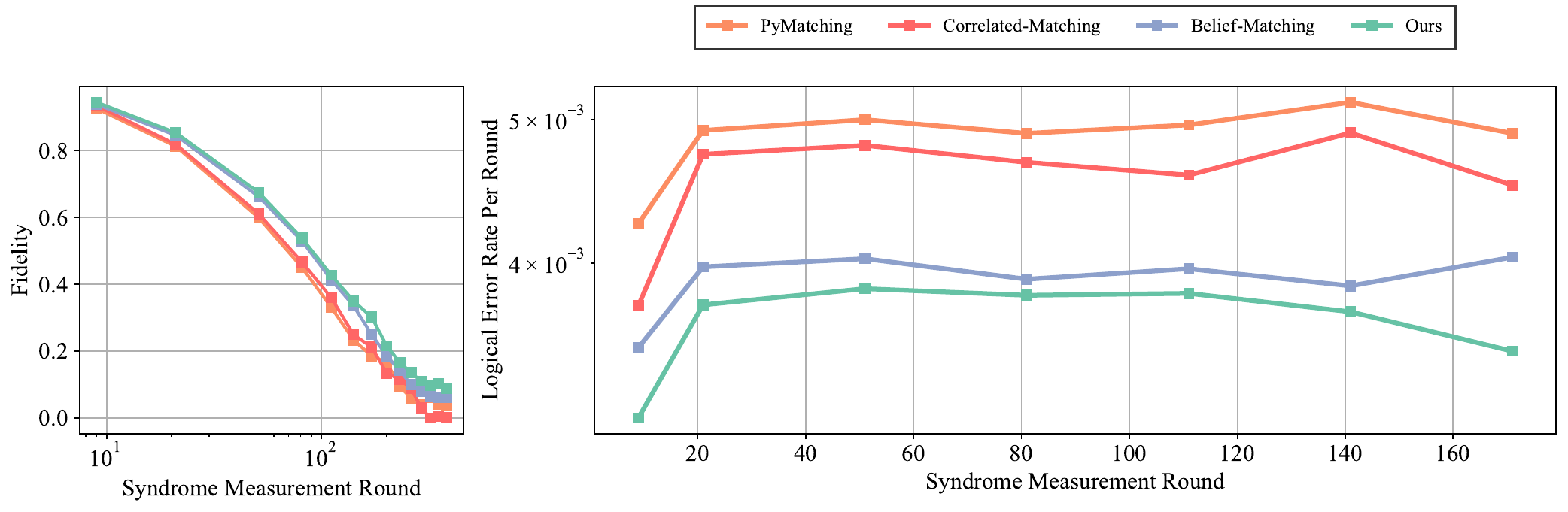}
        \caption{Temporal scalability of the memory experiment with $d = 3$}
        \label{fig:acc_d_3_effp}
    \end{subfigure}
    \begin{subfigure}{\linewidth}
        \includegraphics[width=\linewidth]{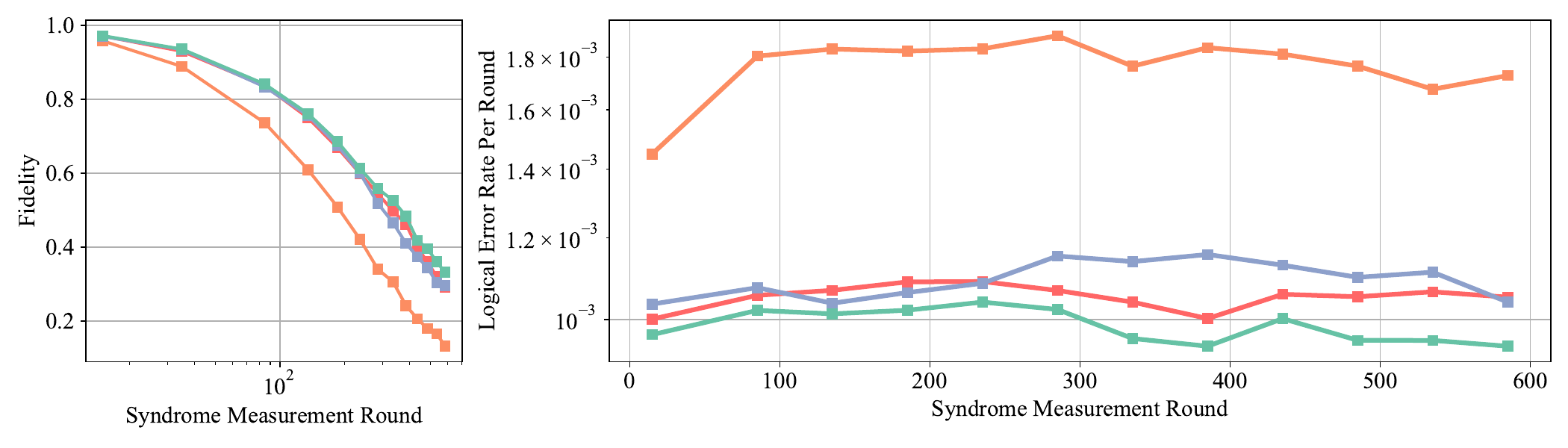}
        \caption{Temporal scalability of the memory experiment with $d = 5$}
        \label{fig:acc_d_5_effp}
    \end{subfigure}
    \begin{subfigure}{\linewidth}
        \includegraphics[width=\linewidth]{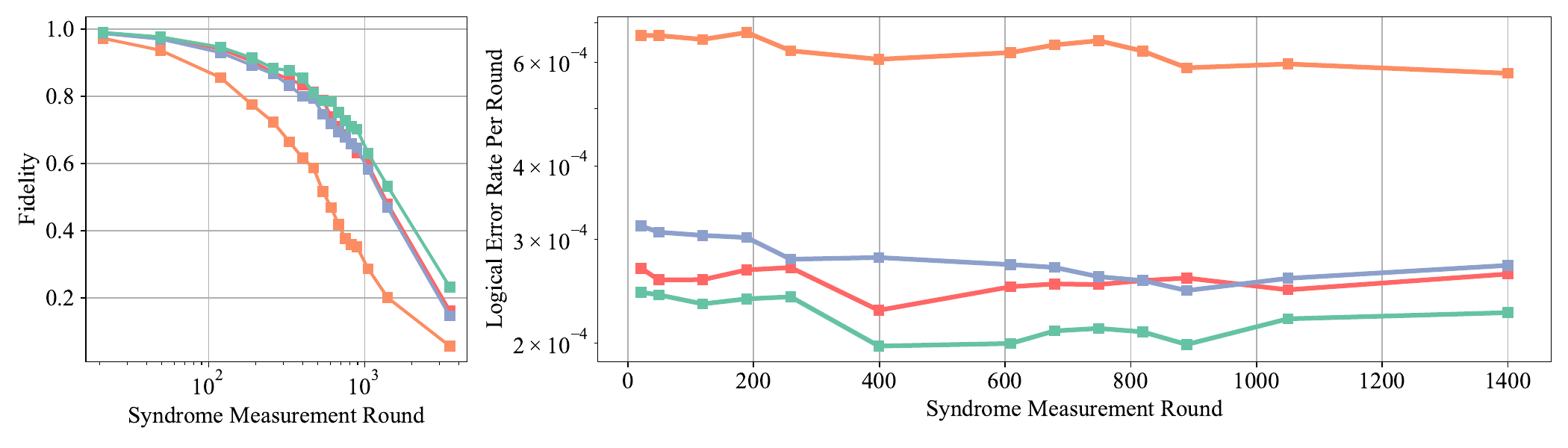}
        \caption{Temporal scalability of the memory experiment with $d = 7$}
        \label{fig:acc_d_7_effp}
    \end{subfigure}
    \caption{Logical error rate per round for memory experiments with increasing number of measurement rounds. The logical error rate per round is plotted for fidelity $F > 0.1$. Each data point is obtained from the logical error rate of at least 10000 shots. While all traditional decoders utilize the global decoding graph for any number of syndrome measurement rounds, our neural network achieves the lowest logical error rates through fully parallel decoding.}
    \label{fig:logical_error_rate}
\end{center}
\end{figure}

Under the realistic yet conservative assumption that the physical error rate is $p=0.003$, we test the temporal scalability of our decoder under this parallel decoding scheme by increasing the number of rounds in memory experiments until the logical error rate reaches $0.5$. We model the dependence of the final logical error rate $p_L$ on the number of rounds $N$ in the same way as~\cite{bausch2024learning}: assume that this dependence can be approximately characterized with the logical error rate per round $\epsilon$, such that every round of the memory experiment has the probability $\epsilon$ of \emph{flipping} the result. The relationship between $p_L$, $\epsilon$, and $N$ is then given by:
\begin{equation}
    p_{L} =\frac{1 - (1-2\epsilon)^N}{2}, \epsilon = \frac{1}{2} (1 - \sqrt[N]{1 - 2 p_L}).
    \label{eq:fitted_ler}
\end{equation}
Note that this implies that $p_L$ is always less than $0.5$. Following~\cite{bausch2024learning}, we also define the \emph{fidelity} $F$ as $1 - 2 p_L$.

We plot the fidelity and the logical error rate per round of all decoders against the number of syndrome measurement rounds in Fig.~\ref{fig:logical_error_rate}, where each data point is averaged over 10,000 shots. The logical error rate per round remains roughly constant over time, suggesting that parallel decoding without local merging does not noticeably harm the accuracy, even as the number of measurement rounds increases. Our decoder significantly outperforms pure MWPM and outperforms Correlated-Matching and Belief-Matching, which is at about the same level as the reported performance of AlphaQubit~\cite{bausch2024learning}. Therefore, it is fair to claim that our parallel decoding scheme maintains state-of-the-art level accuracy when generalizing to any number of syndrome measurement rounds.

\subsubsection{Decoding performance on \textit{Zuchongzhi 3.2}}

\begin{figure}
\begin{center}
    \begin{subfigure}[]{0.5\linewidth}
        \includegraphics[width=\linewidth]{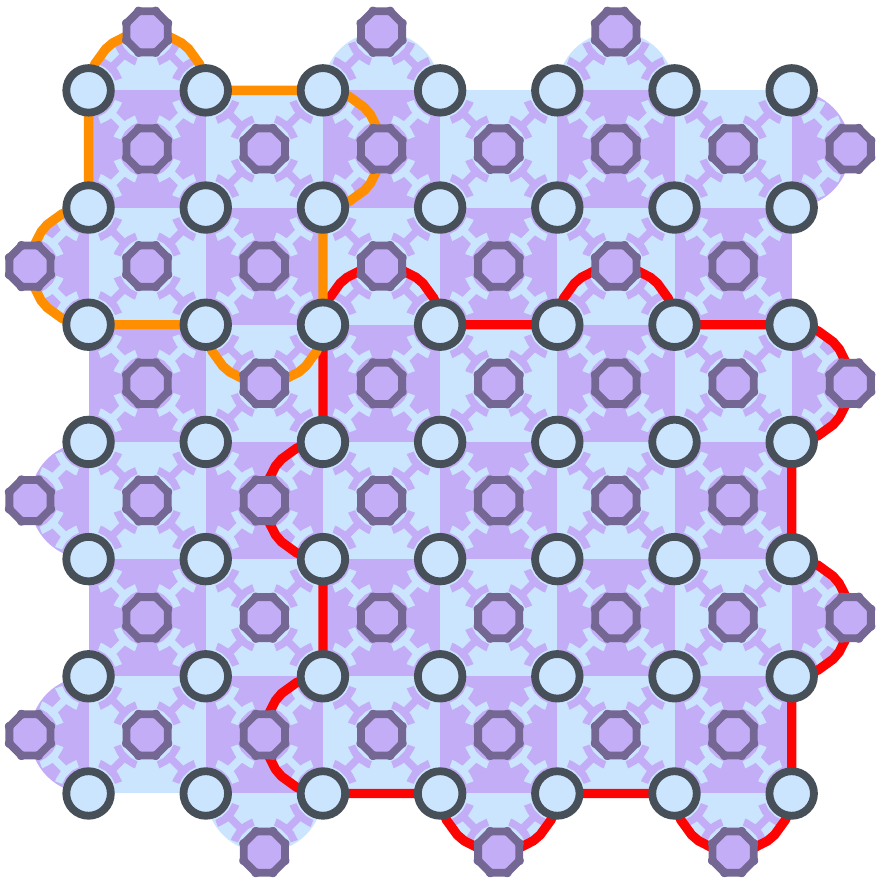}
        \caption{Schematic of the \textit{Zuchongzhi 3.2} quantum processor conducting the decoding experiments}
    \end{subfigure}
    \begin{subfigure}[]{\linewidth}
        \includegraphics[width=\linewidth]{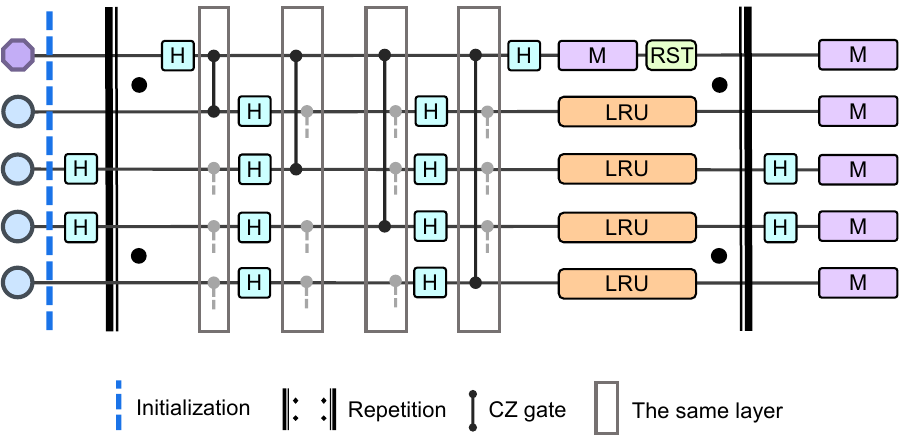}
        \caption{Syndrome extraction circuit}
    \end{subfigure}
    \caption{Quantum processor architecture and the circuits to implement rotated XZZX surface code. (a) Schematic of the quantum processor. Regular octagons represent measurement ancilla qubits, circles represent data qubits, and the H-shaped elements connecting adjacent qubits are tunable couplers. The region outlined in orange is one of the distance-3 patches, and the region outlined in red is one of the distance-5 patches. (b) The circuits to implement rotated XZZX surface code. The Leakage Reduction Unit (LRU) and Reset (RST) are illustrated. The Dynamical Decoupling (DD) pulses are omitted for simplicity.}
    \label{fig:qpu_circuits}
\end{center}
\end{figure}

Beyond circuit-level simulation, our decoder demonstrates high accuracy on decoding experimental data obtained from the \textit{Zuchongzhi 3.2} superconducting quantum processor~\cite{ZCZd7}. Validating the parallel decoding scheme on a physical device is pivotal, as experimental noise environments differ fundamentally from the idealized models used in simulation. While standard depolarizing noise models facilitate large-scale pretraining, physical superconducting processors exhibit complex noise features—including residual crosstalk, leakage, and parameter drift over time. As analyzed in detail in Appendix~\ref{sec:Appendix_E}, these hardware-specific imperfections introduce spatiotemporal correlations that are challenging to capture using standard Pauli error models. This complexity creates a discrepancy between model-based priors and physical reality, necessitating a data-driven approach capable of learning directly from raw hardware events.

\begin{figure}
\begin{center}
    \includegraphics[width=\linewidth]{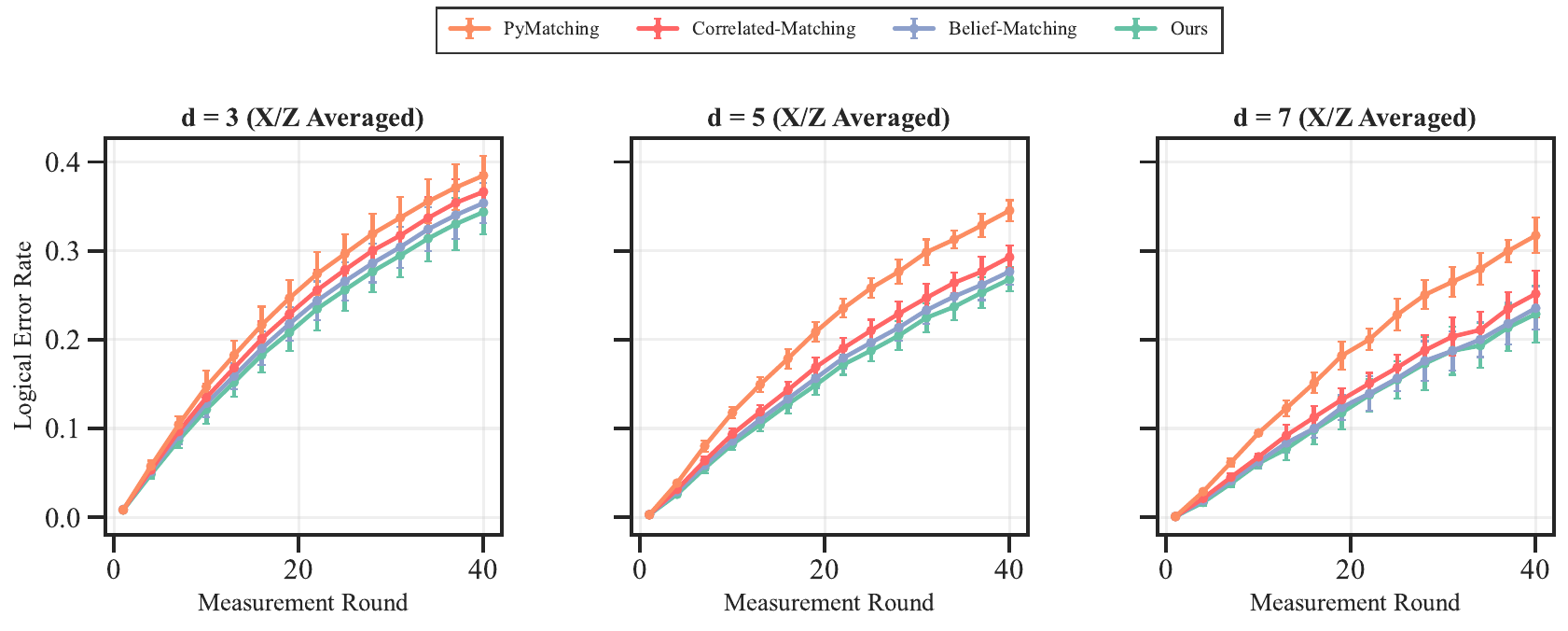}
    \caption{Decoding on \textit{Zuchongzhi 3.2} quantum processor. Compared with traditional global decoding algorithms, our neural decoding scheme achieves the lowest logical error rates across all \emph{odd-indexed} hardware data.}
    \label{fig:logical_error_rate_hardware}
\end{center}
\end{figure}

The processor features a scalable architecture of frequency-tunable transmon qubits arranged in a square lattice, comprising $107$ qubits and $187$ tunable couplers. As shown in Fig.~\ref{fig:qpu_circuits}, for this work, we utilize a selected region of $97$ qubits, configured to implement a distance-7 surface code or multiple independent distance-5 and distance-3 patches. The device is calibrated to support high-performance fault-tolerant protocols. Details about the error rates and the benchmark method are shown in Table~\ref{tab:avg_error_rates}.

\begin{table}[h]
\centering
\caption{Average error rates for single-qubit gates, two-qubit (CZ) gates, and readout in the $97$-qubit region used in this work. }
\label{tab:avg_error_rates}
\begin{tabular}{ccc }
\toprule
Operation & Average error rate (\%) & Benchmark Method \\
\midrule
Single-qubit gates & 0.085 & Parallel single-qubit XEB \\
Two-qubit (CZ) gates & 0.55 & Parallel XEB on qubit pairs\\
Readout & 0.95 & Simultaneous readout\\
\bottomrule
\end{tabular}
\end{table}

We implement the rotated XZZX surface code for its high threshold against biased noise~\cite{XZZX}. The circuit to implement this code is shown in Fig.~\ref{fig:qpu_circuits}. The logical qubit is initialized into an eigenstate of the target logical operator, followed by $N$ rounds of syndrome extraction cycles. Each cycle comprises four layers of entangling CZ gates interspersed with single-qubit gates, scheduled specifically to suppress spatial hook errors. For the final logical readout, we measure all qubits to extract the logical state while simultaneously obtaining syndromes for the final round. Crucially, to maximize the syndrome generation rate, we employ a pipelined sequence at the end of each cycle: fast readout and unconditional active reset are executed on ancilla qubits, occurring concurrently with the leakage reduction unit (LRU) and dynamical decoupling (DD) on data qubits~\cite{ZCZd7}. This optimized timing compresses the total cycle duration to approximately 1 $\mu s$, creating a high-throughput data stream that rigorously tests the real-time processing capability of our decoding architecture.

To systematically evaluate the decoder's scalability, we partition the processor to operate multiple independent patches: nine distance-3 instances, four distance-5 instances, and one distance-7 instance. We apply a comprehensive, device-wide calibration framework to the selected 97-qubit subset to ensure consistent and reliable gate performance across the device. This single calibration is used for all patches (no per-patch retuning), and we execute the same gate and measurement schedule in parallel across patches. For each configuration, logical states are prepared in both the X and Z bases to prevent basis-dependent learning bias. The experiments involve sweeping the number of syndrome extraction cycles from small values to $40$. For every unique circuit configuration, we collect $50,000$ experimental shots. This sample size is selected to ensure sufficient statistical coverage of complex noise features, thereby providing a high-quality dataset for the end-to-end fine-tuning of the neural network.

\begin{table}[htbp]
\centering
\caption{Fitted logical error rate per round on \textit{Zuchongzhi 3.2}.}
\label{tab:ler_comparison}
\begin{tabular}{cccc}
\toprule
Decoding Algorithm & $d=3$ & $d=5$ & $d=7$ \\
\midrule
PyMatching~\cite{higgott2025sparse, Higgott2021PyMatchingAP}  
& $0.01828 \pm 0.00004$ 
& $0.01480 \pm 0.00007$ 
& $0.01278 \pm 0.00012$ \\

Belief-Matching~\cite{higgott2023improved}
& $0.01531 \pm 0.00005$ 
& $0.01023 \pm 0.00004$ 
& $0.00802 \pm 0.00009$ \\

Correlated-Matching~\cite{fowler2013optimal}
& $0.01649 \pm 0.00006$ 
& $0.01115 \pm 0.00004$ 
& $0.00876 \pm 0.00011$ \\

Ours
& $0.01451 \pm 0.00004$ 
& $0.00971 \pm 0.00005$ 
& $0.00783 \pm 0.00010$ \\
\bottomrule
\end{tabular}
\end{table}

We fine-tune our model following the steps in Section~\ref{sec:fine-tuning}, for each basis of the memory experiment and each location of the surface code patch. After fine-tuning, we evaluate the logical error rates across multiple measurement rounds with our parallel decoding scheme. Since the number of measurement rounds $N$ in hardware sampling data is not necessarily an integer multiple of the code distance $d$, we also pad the syndrome data with zeroes so that it can be split into $m = \lceil \frac{N}{d} \rceil$ windows for parallel decoding. For each dataset of $50,000$ experimental shots, we utilize the \emph{even-indexed} subset for DEM calibration and fine-tuning (including DEM-based fine-tuning and end-to-end hardware fine-tuning), while the \emph{odd-indexed} subset is used for evaluation to enable two-fold cross-validation. All traditional decoders use global DEMs specific to patch location, memory basis, and number of measurement rounds. Our neural network models are each fine-tuned on DEM-simulated data and then real hardware data with specific patch location and memory basis, but with mixed numbers of measurement rounds. The logical error rates are averaged over different patch locations and measurement bases, as shown in Fig.~\ref{fig:logical_error_rate_hardware}. We also use the least-squares method to fit the logical error rate per round using the number of measurement rounds $N$ and fidelity $F$ similar to Equation~\ref{eq:fitted_ler}, with a constant factor:
\begin{equation}
C(1-2\epsilon)^N = F = 1 - 2 p_L, \ \ln F = N \ln(1-2\epsilon) + \ln C
\end{equation}
where $C$ denotes the constant factor in the real experiment. The results are listed in Table~\ref{tab:ler_comparison}. Despite the non-ideal nature of hardware noise patterns, our neural decoder attains the lowest logical error rate within a parallel framework, highlighting the effectiveness of end-to-end fine-tuning directly on hardware data.

\subsection{Throughput for real-time decoding}
To evaluate the throughput of our parallel neural network decoder under the real-time decoding scenario, we benchmark the throughput performance on the TPU v6e. The decoding throughput is measured as the decoding latency per syndrome extraction round. As mentioned before, for the superconducting quantum processor, the decoding latency per round should be below $1 \mu s$ to achieve sufficient throughput for real-time decoding~\cite{Tan2022ScalableSD, Skoric2022ParallelWD, google2025quantum}, which means that no syndrome backlog occurs and the reaction time is kept constant regardless of the syndrome extraction rounds. We test the throughput using memory experiments that last for up to $N \sim 10^5$ measurement rounds. Since our decoding scheme is fully parallel, we can split the arbitrarily long decoding graph into independent windows (each window contains $n = 3 d$ rounds) and decode them in parallel. The throughput is evaluated by:
\begin{equation}
    \text{decoding time per round} = \frac{\text{decoding time}}{N} = \frac{\text{decoding time}}{m \times d}
\end{equation}
where $m$ is the number of independent decoding windows. The throughput of our parallel decoding scheme compared with AlphaQubit is listed in Table~\ref{tab:throughput}.

\begin{table}[htbp]
\centering
\caption{Decoding throughput in terms of decoding time per round. For each code distance $d$, we leverage the maximum achievable parallelism on a single TPU v6e.}
\label{tab:throughput}
\begin{tabular}{ccc}
\toprule
Code distance & AlphaQubit (estimated from~\cite{han2015learning}) ($\mu s$) & Ours ($\mu s$) \\
\midrule
3  & $19$ & $0.781 \pm 0.041$ \\
5  & $25$ & $0.788 \pm 0.036$ \\
7  & $30$ & $0.789 \pm 0.031$ \\
9  & $38$ & $0.798 \pm 0.030$ \\
11 & $48$ & $0.812 \pm 0.013$ \\
13 & $55$ & $0.815 \pm 0.019$ \\
15 & $65$ & $0.837 \pm 0.021$ \\
17 & $85$ & $0.841 \pm 0.023$ \\
19 & $100$ & $0.896 \pm 0.018$ \\
21 & $105$ & $0.921 \pm 0.015$ \\
23 & $130$ & $0.932 \pm 0.015$ \\
25 & $170$ & $0.979 \pm 0.021$ \\
27 & ---   & $1.002 \pm 0.017$ \\
\bottomrule
\end{tabular}
\end{table}

The throughput demonstrates two key points. First, by adopting a fully parallel scheme, real-time decoding for arbitrarily long quantum memory experiments can be achieved by stacking sufficient parallel resources, thereby ensuring no syndrome backlog. Fortunately, modern neural network inference hardware, such as TPU and GPU, naturally provides the required level of parallelism: a single TPU v6e is sufficient to support our parallel real-time decoding for code distances up to $d = 25$. Second, our approach exhibits better scalability with respect to increasing code distance, beyond that of the recent independent work---Google's AlphaQubit2~\cite{senior2025scalable}, which only supports sufficient throughput for code distance up to $d = 11$ (AQ2-RT).

\subsection{Scalability}
In this work, we trained our model on surface codes up to distance $7$ to validate the feasibility of real-time neural decoding for superconducting qubits. Nevertheless, realizing large-scale fault-tolerant quantum computation will require neural decoders that scale efficiently---a critical consideration for future development. In the previous subsection on throughput, we demonstrated that our neural decoder can support real-time decoding for surface codes up to distance 25 on a single TPU v6e, highlighting its superior scalability. However, another critical consideration is the training cost.

The training overhead of our models is summarized in Table~\ref{tab:training_overhead}. As the code distance increases, the required number of training samples grows rapidly. Our training was conducted on NVIDIA GeForce RTX 4090 GPUs. This is a setup we can reliably use for long-duration training, yet it remains modest even by academic standards. Compared to the original AlphaQubit, which demonstrated successful training up to distance $d=11$, our network architecture introduces no additional fundamental barriers to scaling. We therefore expect that training up to $d=11$ can be achieved with comparable hardware resources and is likely extendable to somewhat larger code distances with modest improvements, as suggested in~\cite{bausch2024learning}.

A few months after our initial submission to arXiv---and during the preparation of this revised manuscript, which now includes experimental results---Google released AlphaQubit2 (AQ2)~\cite{senior2025scalable}, a follow-up to AlphaQubit (AQ1). AQ2 introduces an enhanced neural architecture that enables successful training up to code distance $d=23$. Crucially, this advance stems from a shift in training strategy, as suggested in~\cite{bausch2024learning}: whereas AQ1 was trained separately for each individual code distance, AQ2 is trained jointly across multiple code distances. This multi-distance training paradigm significantly improves generalizability and scalability, allowing the model to handle larger codes without retraining from scratch. In principle, we see no fundamental barriers to adopting a similar strategy in our framework. However, to better reflect the parallel and independent nature of our work relative to AQ2, we defer such an extension to future work.

Regarding other distinctions with AQ2: by decoupling the transformer and the RNN, AQ2 introduces a degree of intra-decoding-round parallelism. Specifically, multiple rounds of stabilizer measurements are compressed into a single temporal layer and processed simultaneously by the transformer---without waiting for outcomes from prior rounds. The RNN then acts as a ``merge'' module, aggregating the intermediate transformer outputs to produce the final logical decoding result.

AQ2 demonstrates that neural decoders can scale to larger code distances when equipped with larger models,  more carefully designed training strategies, and more substantial computational resources. In contrast, our framework achieves real-time decoding through \textit{fully} parallel processing of disjoint subgraphs, offering a high degree of intrinsic parallelism that is independent of the underlying neural architecture. As such, it constitutes a general-purpose decoding strategy that can be seamlessly integrated with more advanced neural decoders---including AQ2---in future implementations.

\begin{table}[ht]
\centering
\caption{Training overhead.}
\label{tab:training_overhead}
\begin{tabular}{c cc cc}
\toprule
\multirow{2}{*}{Code distance} 
& \multicolumn{2}{c}{Pretraining} 
& \multicolumn{2}{c}{Fine-tuning} \\
\cmidrule(lr){2-3} \cmidrule(lr){4-5}
& Samples 
& GPU hours$^{\dagger}$ 
& DEM-based samples$^*$ 
& End-to-end samples$^*$ \\
\midrule
3 & 90 M  & 50   & 10 M   & 25{,}000 \\
5 & 300 M & 400  & 20 M   & 25{,}000 \\
7 & 1 B   & 2000 & 100 M  & 25{,}000 \\
\bottomrule
\end{tabular}
\begin{flushleft}
\footnotesize
$^{\dagger}$: Training times reported are based on NVIDIA GeForce RTX 4090 GPUs.\\
$^*$: For each configuration of patch location and memory basis.
\end{flushleft}
\end{table}

\section{Summary and Outlook}

Most existing learning-based decoders, despite achieving high accuracy, suffer from high latency, limited throughput, and poor scalability, rendering them unsuitable for the real-time demands of fault-tolerant quantum computing. AlphaQubit, a leading example of such decoders, employs an RNN-based architecture that enables accurate decoding of quantum memory experiments at larger scales and across an arbitrary number of rounds, all while maintaining relatively low computational overhead. However, its inherently sequential design precludes parallel inference, resulting in throughput that falls one to two orders of magnitude short of real-time requirements.

In this work, we resolve the throughput bottleneck, which is the primary obstacle limiting the applicability of AlphaQubit-type decoders to large-scale fault-tolerant quantum computing. Without altering the underlying AlphaQubit architecture, we introduce a novel scheme that preserves the decoder's high accuracy while enabling parallelized inference, thereby substantially boosting decoding throughput. Our approach makes neural decoders viable for real-time fault-tolerant quantum computing, enhancing the robustness and reliability of the decoding process. To validate its superior accuracy, we trained and evaluated our model on surface codes with distances up to $7$, the largest surface code instances demonstrated to date on superconducting qubits.

In addition, our simulations convincingly demonstrate that our approach can enable scalable, real-time decoding of quantum memory experiments at a scale beyond previous reach: reported data indicate that a single TPU v6e, which is not even the latest generation, can enable real-time decoding for surface codes up to distance $25$, typically considered adequate for practical fault-tolerant quantum applications. However, this marks only a starting point; much work remains to advance learning-based decoders like AlphaQubit toward practical feasibility in fault-tolerant quantum computing, including the following directions:

\textbf{Hardware deployments.}
Our model has not been specifically optimized for deployment on current hardware. While it provides sufficient throughput to handle an arbitrary number of syndrome measurement rounds, the final constant decoding latency---reaction time, has not been tuned for small-scale experiments. Consequently, existing real-time decoding architectures based on traditional decoders---such as those in~\cite{google2025quantum, zhang2025latte, wu2023fusion, wu2025micro}, remain strong competitors in near-term settings.

Nevertheless, our approach offers a general and flexible framework that is not inherently tied to the original AlphaQubit architecture. From a systems perspective, the proposed design is largely orthogonal to the choice of the underlying model: it can be seamlessly integrated with more efficient sequence models, including state-space representations~\cite{gu2021efficiently, gu2024mamba}. Moreover, the framework is compatible with standard inference-time optimizations---such as quantization and pruning~\cite{jacob2018quantization, han2015learning}---as well as execution-level techniques like KV-cache reuse~\cite{dao2022flashattention}, all of which improve computational and memory efficiency without altering model semantics.

\textbf{Beyond memory experiments.} 
We outline how our parallel neural decoder may be generalized to decode lattice surgery logical operations in a relatively straightforward way. As noted in~\cite{bombin2023modular}, the decoding graph of a lattice surgery procedure is modular. Concretely, we can cut the decoding graph to $d \times d \times d$ pieces, let each piece be a core region, and add all vertices within distance $d$ of the core region as the buffer region to build the decoding windows. Since the decoding graph is built up from a discrete set of $d \times d \times d$ building blocks, the number of distinct types of windows that could be built this way is in principle constant, and it should be possible to train a neural decoder to handle them. The decoding of individual windows remains fully parallel, and thus with an amount of computational resources proportional to the spatial footprint of the lattice surgery procedure, the throughput requirement can be satisfied. We note that this should not affect the asymptotic hardware complexity of the entire quantum computer, since the number of physical qubits needed is already proportional to the spatial footprint. Thus, the proposed parallel decoding scheme enables lattice surgery with computational resources that scale linearly with the number of logical qubits, making it practically feasible for hardware deployment.

Looking ahead, we anticipate that learning-based approaches---powered by the parallel decoding framework developed in this work and ongoing advances---will serve as a cornerstone of quantum computing infrastructure, ultimately becoming integral to quantum computers themselves and enabling more robust and cost-efficient quantum memory and operations.

\backmatter

\bmhead{Acknowledgements}

The work is supported by National Key Research and Development
Program of China (Grant No.\ 2023YFA1009403), Quantum Science and Technology-National Science and Technology Major Project (Grant No.\ 2021ZD0300200), National Natural Science
Foundation of China (Grant No.\ 12347104), Beijing Natural Science Foundation
(Grant No.\ Z220002), CCF-QuantumCtek Superconducting Quantum Computing Special Cooperation Program (No.\ CCF-QC2025001), the Major Key Project of PCL (Grant No.\ PCL2024A06 and No.\ PCL2025A10), the Shenzhen Science and Technology Program (Grant No.\  RCJC20231211085918010), Zhongguancun Laboratory, Tsinghua University, and Pengcheng Laboratory.

\newpage

\begin{appendices}

\section{Model Architecture in Detail}
\label{sec:Appendix_B}

Table~\ref{tab:model_parameter} shows the model parameters in detail. The key modules and the design considerations are described in the following sections.

\begin{table}
\centering
\caption{Module parameters setting.}
\small
\label{tab:model_parameter}
\begin{tabular}{c c c}
\toprule
\textbf{Module} & \textbf{Layer} & \textbf{Parameters} \\
\midrule
\multirow{2}{*}{\textbf{StabilizerEmbedder}} 
    & Linear & $\text{Linear}(1 \rightarrow d_{\text{model}})$ \\
    & ResNet & 2 $\times$ Conv1d($d_{\text{model}}, d_{\text{model}}$) \\
\midrule
\multirow{3}{*}{\textbf{SyndromeTransformer}} 
    & Multi-Head Attention & $d_{\text{model}}, n_{\text{head}}$ heads \\
    & Feedforward & $d_{\text{model}} \rightarrow 5d_{\text{model}} \rightarrow d_{\text{model}}$ \\
    & Dilated Convolutions & 3 $\times$ Conv2d($d_{\text{model}}, d_{\text{model}}$)\\
\midrule
\multirow{1}{*}{\textbf{RNNCore}} 
    & Transformer Layers & 3 $\times$ \textbf{Syndrome Transformer} \\
\midrule
   \multirow{3}{*}{\textbf{Readout}} & 2D Convolution & 1 $\times$ Conv2d ($d_{\text{model}},  d_{\text{model}}$)\\
    & Logical Pooling & ($d \times d \to d$) \\
    & Linear & Linear($d \cdot d_{\text{model}}  \rightarrow 1$)\\
\bottomrule
\end{tabular}
\end{table}

\subsection{Syndrome embedding}
\subsubsection{Boundary padding strategy}

\label{sec:padding}
\begin{figure}
\begin{center}
\includegraphics[width=\linewidth]{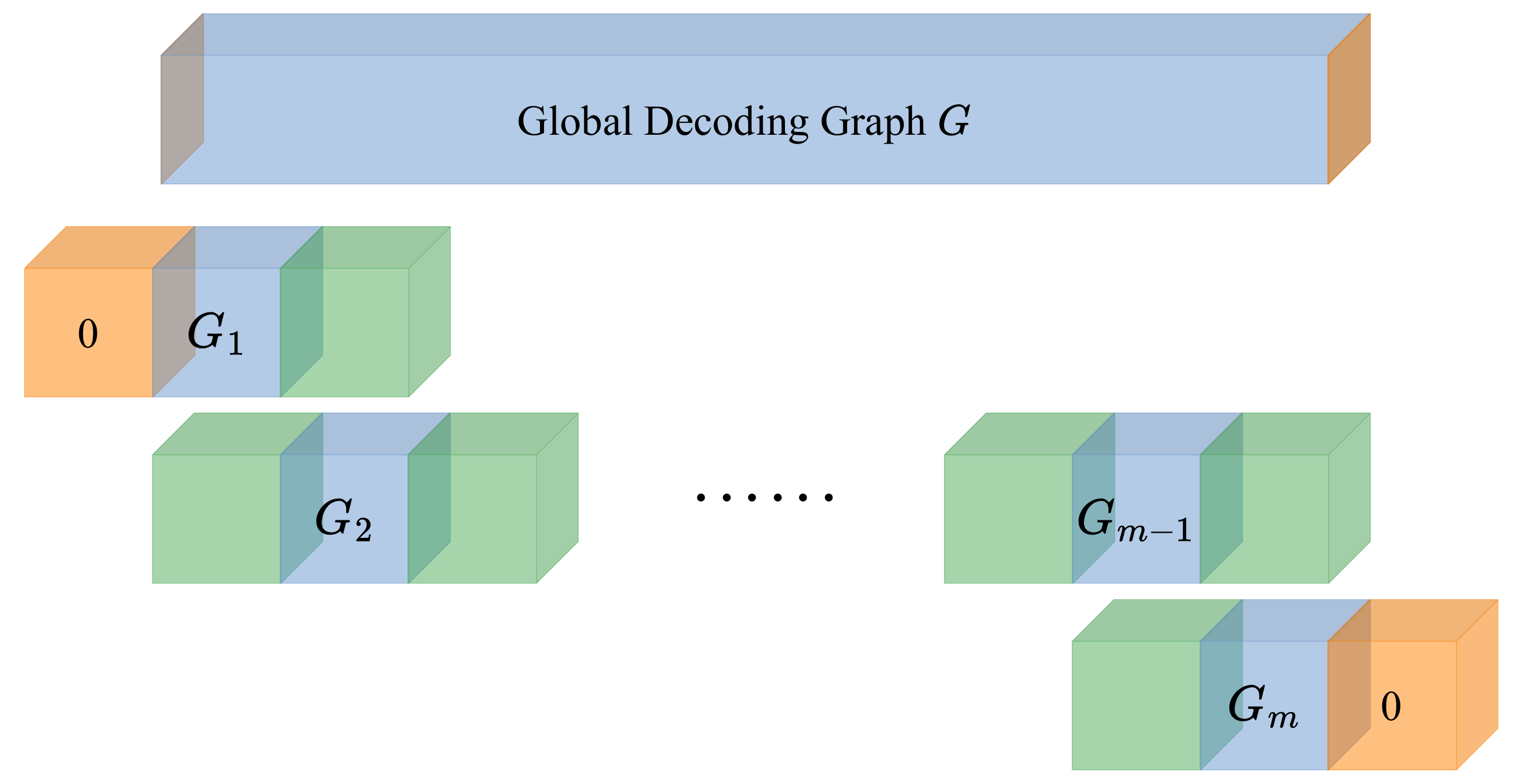}
\caption{Illustration for the global decoding graph and three kinds of decoding windows in the quantum memory experiment. Suppose the global syndrome graph can be partitioned into $m$ decoding windows, then the decoding windows can be divided into $1$ \emph{initial window}, $1$ \emph{final window} and $(m-2)$ \emph{bulk window}s.}
\label{fig:parallel_window_decoding}
\end{center}
\end{figure}

In the memory experiment, the first and the final syndrome measurement rounds are closed time boundaries as described in~\cite{Tan2022ScalableSD}. This means that neither a single $Z$ detector flip in the first round nor one in the final round cannot be interpreted as an ancilla measurement error, since the logical qubit is initialized by preparing all \emph{data} qubits to $\ket{0}$ and measured by measuring all \emph{data} qubits in the $Z$ basis. However, when dividing a long decoding graph into separate decoding windows, intermediate decoding windows will have ``open time boundaries'', and thus require buffer regions for accurate decoding. Therefore, there are three different types of decoding windows as depicted in Fig.~\ref{fig:parallel_window_decoding}: the \emph{initial window}, the \emph{bulk window}s, and the \emph{final window}.

In works such as~\cite{Tan2022ScalableSD}, the total number of syndrome rounds in each window is kept constant (as much as possible). This means that the initial window and the final window have longer core regions since they each only have one buffer region instead of two. We instead opt to align the semantics of all three types of windows by keeping the size of the \emph{core regions} constant. For the initial window and the final window, we add boundary padding (orange regions in Fig.~\ref{fig:parallel_window_decoding}) filled with zeros in place of the nonexistent buffer regions. Since a zero-filled padding naturally suppresses matching across the time boundary, we no longer need to define special semantics for those ``closed time boundaries''. This ensures that the three window types have consistent semantics and shapes, simplifying our neural network design while improving load balancing in parallel decoding.

Instead of applying specialized encoding to the three different window types, we align their semantics in the data, enabling a single network to handle different window types and thereby improving the training efficiency. Moreover, joint learning across the three window types further enhances the self-consistency.

\subsubsection{Lattice-based syndrome encoding}

We map the locations of all stabilizers in a distance-$d$ surface code into a $(d + 1) \times (d + 1)$ array, as shown in Fig.~\ref{fig:overview}. Each syndrome extraction cycle can then be naturally embedded into a $(d + 1) \times (d + 1)$ tensor $\mathbf{E}$, with $1$ representing a syndrome defect and $0$ for the absence of a defect. For locations where no syndrome vertices exist, we also use 0 as a natural padding value. We prepare our data for use in the transformer by flattening each $(d + 1) \times (d + 1)$ tensor into a vector. An entire decoding window is thus encoded into a sequence of $b+c+b$ vectors, where $b$ is the buffer size and $c$ is the core region size. The encoding process can be excuted in a streaming manner.

\subsubsection{Syndrome convolution}\label{sec:conv1d}

We move our simplified ResNet layer, which only contains \texttt{Conv1D} with residual connection~\cite{He2015DeepRL}, to the front of the positional encoding (\textbf{PE}). This modification is motivated by our observations: since convolutional layers do not inherently require \textbf{PE}, we believe that applying \texttt{Conv1D} directly to the flattened syndrome data can better capture local error features, without the loss of original information that \textbf{PE} might introduce. The output of the ResNet is a sequence of tensors $\mathbf{E}_\text{convolved}$ with shape $(d + 1)^2 \times d_\text{model}$.

\subsubsection{Positional encoding}
\label{sec:positional_encoding}

To help the transformer module capture spatial relationships, the encoder augments the convolutional representation with a deterministic positional encoding~\cite{vaswani2017attention}. A fixed sinusoidal encoding \(\mathbf{PE}(t)\in\mathbb{R}^{d_\text{model}}\) is generated as:

\begin{equation}
\mathbf{PE}_{2i}(t) = \sin \Bigl(\dfrac{t}{10000^{2i/d_\text{model}}}\Bigr), \ \mathbf{PE}_{2i+1}(t) =\cos \Bigl(\dfrac{t}{10000^{2i/d_\text{model}}}\Bigr)
\end{equation}
where $i = 0,1,\dots,\left\lfloor d_\text{model}/2\right\rfloor$ and $t = 0,1,\dots,(d+1)^2-1$. This yields a spectrum of $d_{model}$ sinusoid sequences with length $(d+1)^2$, whose wavelengths form a geometric progression. The final stabilizer embedding $\mathbf{S}$ is obtained by the element-wise summation:
\begin{equation}
\mathbf{S} = \mathbf{E}_\text{convolved} + \mathbf{PE}.
\end{equation}
This embedding endows the model with translation-equivariant yet position-sensitive features, capturing local error patterns while preserving global location information.

\subsection{RNN}
The RNN receives the complete decoding window syndrome data after undergoing the embedding layer, including the left buffer region, the core region, and the right buffer region, one round at a time in temporal order. The initial decoder state $\mathbf{D_0}$ is set to be zero for all kinds of decoding windows. At round $n$, the RNN core combines the input stabilizer embedding tensor  $\mathbf{S}_{n}$ with the current decoder state $\mathbf{D_{n-1}}$, normalized by $\sqrt{2}/{2}$. This combined input is then processed through three sequential syndrome transformer layers, and the output becomes the new decoder state:
\begin{equation}
    \mathbf{D_n} = \texttt{RNNCore}((\mathbf{S_{n}} + \mathbf{D_{n-1}}) \times 0.707).
\end{equation}

The syndrome transformer module inside the RNN is primarily designed to capture more global error information, with the \texttt{self-attention}~\cite{vaswani2017attention} as its key component. We also follow the implementation in~\cite{bausch2024learning}, where the output after attention is scattered onto a 2D grid for \texttt{dilated convolution}, aggregating error information further over the surface code topology. Both the input and the final output of the syndrome transformer module has the same shape as the embedded syndrome input.

\subsection{Readout}
After processing all rounds of a decoding window, the final decoder state $\mathbf{D_{n}} = \mathbf{D_{b+c+b}}$ is passed to the \emph{readout} module. The readout module first shrinks the dimension from $(d + 1) \times (d + 1)$ to $d \times d$ using a \texttt{Conv2D} layer without padding. Then, the $d \times d$ grid is pooled only in the direction perpendicular to the logical operator, and the resulting $d \times d_\text{model}$ feature tensor is projected to a singular logit via a simple linear layer as the final logical prediction $\hat{y}^c_i$ for the core region. The entire process, from the syndrome input to the logical readout, is also comprehensively outlined in Algorithm~\ref{alg:rnn_decoder_alg}, providing a clearer and more structured overview.

\section{Training Details}
\label{sec:Appendix_C}
\subsection{Pretraining stage}
During the pretraining stage, all training data are generated randomly using Stim~\cite{Gidney2021StimAF} by two steps. First, we generate the global syndrome data with about $3d\sim5d$ syndrome measurement rounds. Second, we split the global syndrome data into $3\sim5$ decoding windows, and mix them together as our training dataset. In our training process, we generally adopt a larger physical error rate such as $p \sim 0.005$ at the beginning, which enables the model to more quickly learn difficult decoding scenarios and facilitate convergence. Nevertheless, we also observe that training the model under varying physical error rates can yield improved performance.

The hyperparameters, model size are presented in Table~\ref{tab:training_gpu_hours}. During the training process, we observed behavior consistent with a previously reported finding~\cite{bausch2023learning}---that the number of training samples needed to match the performance of traditional algorithms~\cite{higgott2025sparse, Higgott2021PyMatchingAP} grows exponentially with code distance. We further discovered an additional key trend: as the testing physical error rate $p$ decreases, the required number of training samples also increases exponentially. 

\begin{table}[ht]
\centering
\caption{Hyperparameter setup for different code distances. Hyperparameters include the feature dimension $d_\text{model}$, the number of attention heads $n_\text{head}$, the batch size $B$, and the learning rate $\alpha$.}
\label{tab:training_gpu_hours}
\begin{tabular}{|c|c|c|c|c|c|c|c|}
\hline
\multirow{2}{*}{Code distance} & \multicolumn{6}{c|}{Hyperparameters} & \multirow{2}{*}{Model size$^*$} \\ 
\cline{2-7} & $d_{\text{model}}$ & $n_\text{head}$ & $B_\text{init}$ & $B_\text{final}$ & $\alpha_\text{init}$ & $\alpha_\text{final}$ & \\
\hline
3 & 192 & 8 & 256 & 512 & 1e-4 & 5e-5 & 4,586,113    \\
5 & 192 & 8 & 576 & 576  & 1e-4 & 5e-5 & 4,586,497 \\
7 & 192 & 8 & 448 & 1344  & 5e-4 & 5e-5 & 4,586,881 \\
\hline
\end{tabular}
\begin{flushleft}
\footnotesize
$^*$ The difference in model size arises from slight variations in the parameters of the linear projection network in the Readout module corresponding to different code distances.
\end{flushleft}
\end{table}

\subsection{Fine-tuning stage}
During the hardware fine-tuning stage, we first calibrate the hardware DEM using the \emph{even-indexed} subset of the hardware data, i.e., 25000 samples for each patch position and X/Z basis by the correlation analysis method~\cite{chen2022calibrated, google2021exponential, Acharya2022SuppressingQE}. The DEM is constructed based on the open-source implementation in~\cite{zhangcorrelation}, by fitting to syndrome data so that the resulting sampling matches the empirical distributional characteristics, including marginal distributions and higher-order joint expectations.

After calibration, we sample the training data from the hardware DEM to feed the fine-tuning of our neural network. All sampled data are segmented into different kinds of windows for mixed training. In the end, the \emph{even-indexed} subset data and the corresponding logical observables are also used to perform the end-to-end global fine-tuning with the Soft-XOR method.

\section{Ablation Study}

While AlphaQubit~\cite{bausch2023learning, bausch2024learning} has provided a comprehensive collection of ablation studies of different modules in the model design, we are focusing on several critical components in our modifications and designs. Since training a model from scratch entails considerable computational resources and time costs, which grow almost exponentially with the code distance $d$, we perform the ablation studies only on the $d = 3$ model.

The following model variations are trained and evaluated:
\begin{itemize}
    \item \textbf{Baseline}: The baseline model corresponding to Fig.~\ref{fig:neural_network_architecture} and Table~\ref{tab:model_parameter}.
    \item \textbf{PEARes}: Positional encoding ahead of the ResNet module.
    \item \textbf{BNTRST}: In batch normalization, setting \texttt{track\_running\_stat} to True.
    \item \textbf{GatedDenseBlock}: Keeping AlphaQubit's \texttt{GatedDenseBlock} module rather than the \texttt{Feedforward} layer directly from~\cite{vaswani2017attention}.
    \item \textbf{SepWindow}: Utilizing independent trained models rather than one model for decoding, to validate the effectiveness of self-consistency.
    \item \textbf{NoMixTrain}: Training process is designed to train three kinds of window data in different stages separately, each with $1/3$ training samples.
    \item \textbf{NoRec}: Training without the recurrent training method in Section~\ref{sec:recurrent_training}.
\end{itemize}

\begin{figure}[htbp]
\begin{center}
\includegraphics[width=0.8\linewidth]{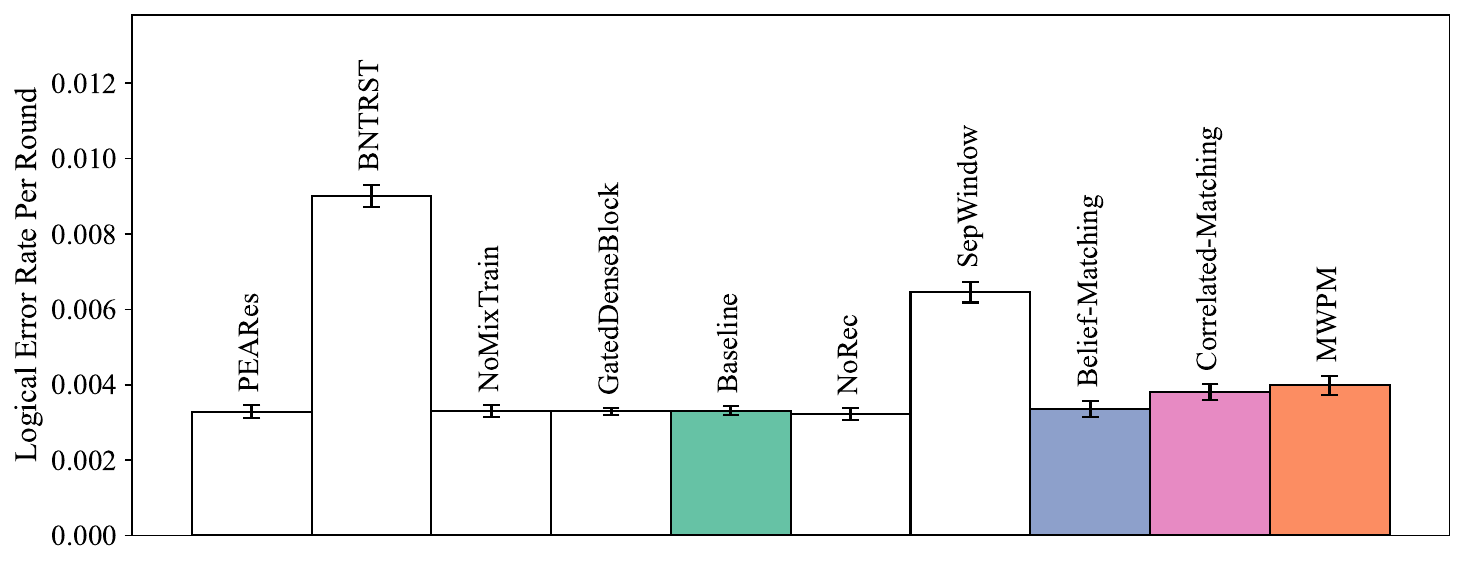}
\caption{Logical error rate per round of different model variations for $d = 3$, along with traditional decoding algorithms.}
\label{fig:ablation_study}
\end{center}
\end{figure}

To ensure a fair comparison, all models were trained across three datasets of $200M$ samples each (corresponding to $p = 0.006$, $p = 0.005$, and $p = 0.004$). After training, we conduct the memory experiment with $3d$ rounds, i.e., divide into three decoding windows and decode them in parallel, under $p = 0.003$ to evaluate the decoding accuracy. Each variation's logical error rate per round is illustrated in Fig.~\ref{fig:ablation_study}. 

We summarize our main observations from the ablation study as follows:

\begin{enumerate}
    \item The order between \textbf{PE} and ResNet appears to be not so critical. 
    \item Disabling \texttt{track\_running\_stat} in BatchNorm is essential, likely because batches seen by the BatchNorm layer in training are not identically distributed due to the RNN architecture. A possible explanation is that our input representation requires the RNN to ``remember'' the current round number, which causes this information to affect the decoder state. Therefore, enabling \texttt{track\_running\_stat} makes the network behave too differently in evaluation from in training.
    \item The choice between the \texttt{GatedDenseBlock} and the \texttt{Feedforward} layer has only a minor impact overall. Nonetheless, we observe in our training that the \texttt{Feedforward} layer achieves better convergence for larger code distances.
    \item If different models are trained completely independently for parallel decoding, then even though each single model has converged, the global accuracy after parallel decoding remains low due to the lack of self-coordination. 
    \item Staging the training by window types or using the mixed samples directly has little influence on the final model capacity. 
    \item Although recurrent training is not so critical for the final accuracy, we observe that models trained with multi-layer recurrent prediction tend to converge much faster than the singular prediction mode. 
\end{enumerate}

\section{Noise Analysis of Hardware Data}
\label{sec:Appendix_E}
Unlike circuit-level simulations, experimental superconducting processors exhibit noise mechanisms that are intrinsically structured in space and time. In particular, non-Pauli processes such as leakage can be long-lived and can induce correlated error patterns that persist across multiple QEC cycles, thereby violating the i.i.d.\ assumptions commonly used in circuit-level noise model. This motivates an \emph{in-situ} diagnosis of the hardware error environment directly from detection events.

\begin{figure}[t]
\centering
\begin{subfigure}[]{\linewidth}
    \centering
    \includegraphics[width=\linewidth]{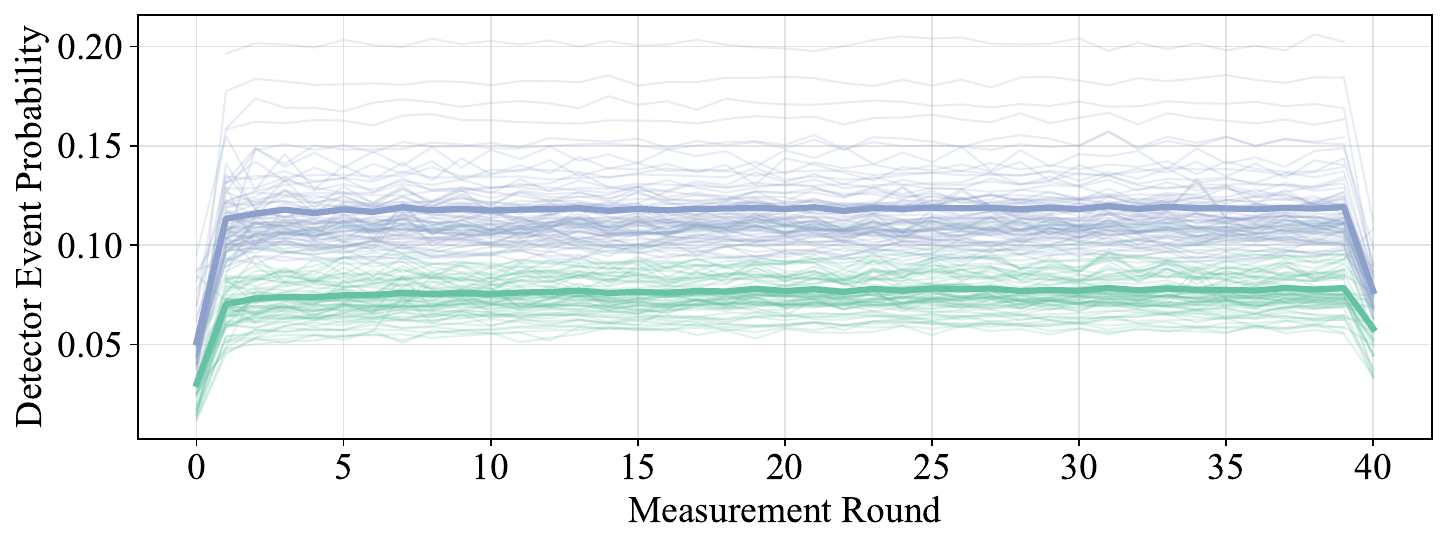}
    \caption{DEP visualization of $d = 3$}
    \label{fig:def:a}
\end{subfigure}
\begin{subfigure}[]{\linewidth}
    \centering
    \includegraphics[width=\linewidth]{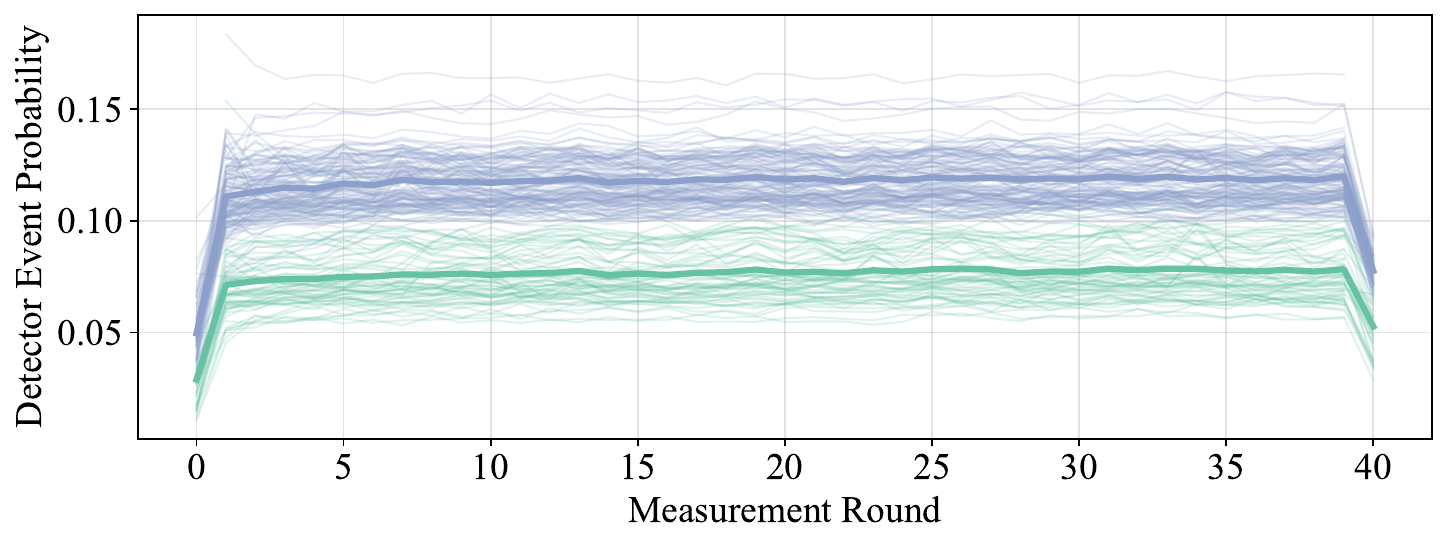}
    \caption{DEP visualization of $d = 5$}
    \label{fig:def:b}
\end{subfigure}
\begin{subfigure}[]{\linewidth}
    \centering
    \includegraphics[width=\linewidth]{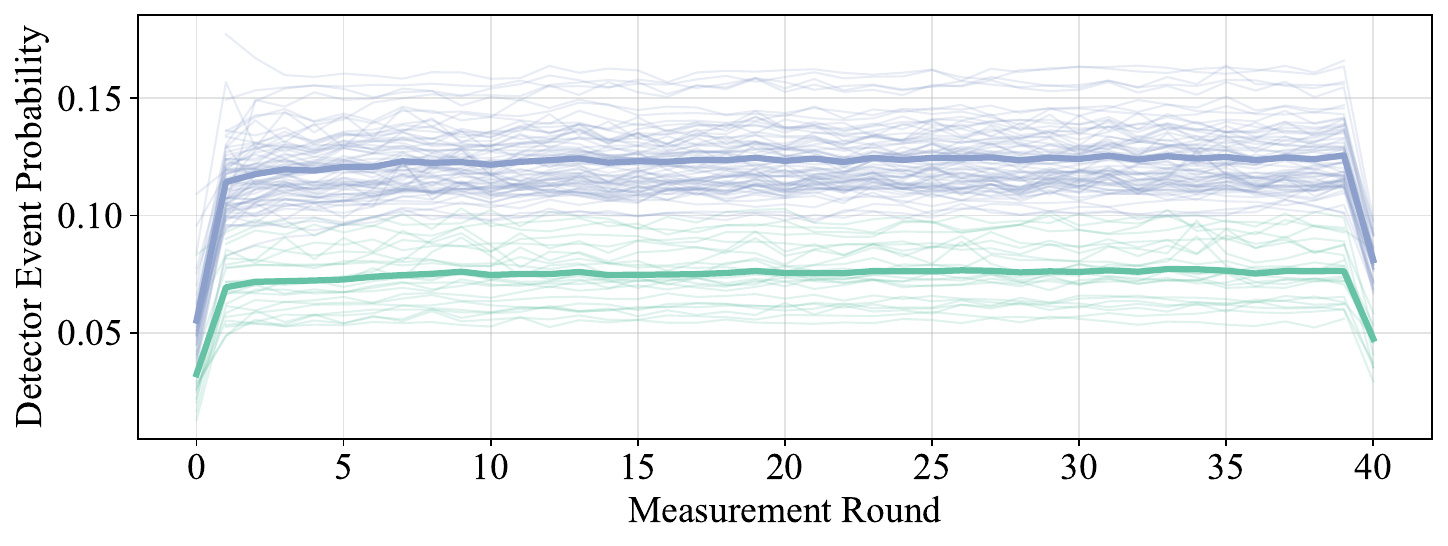}
    \caption{DEP visualization of $d = 7$}
    \label{fig:def:c}
\end{subfigure}

\caption{Detector event probability of \textit{Zuchongzhi 3.2} for two detector classes: weight-2 (blue) and weight-4 (orange). The transparent curves show the DEP of each detector. The solid curves show the mean probability of all detectors with the same weight.}
\label{fig:def}
\end{figure}

\begin{figure}
\centering
\begin{subfigure}[]{0.5\linewidth}
    \centering
    \includegraphics[width=\linewidth]{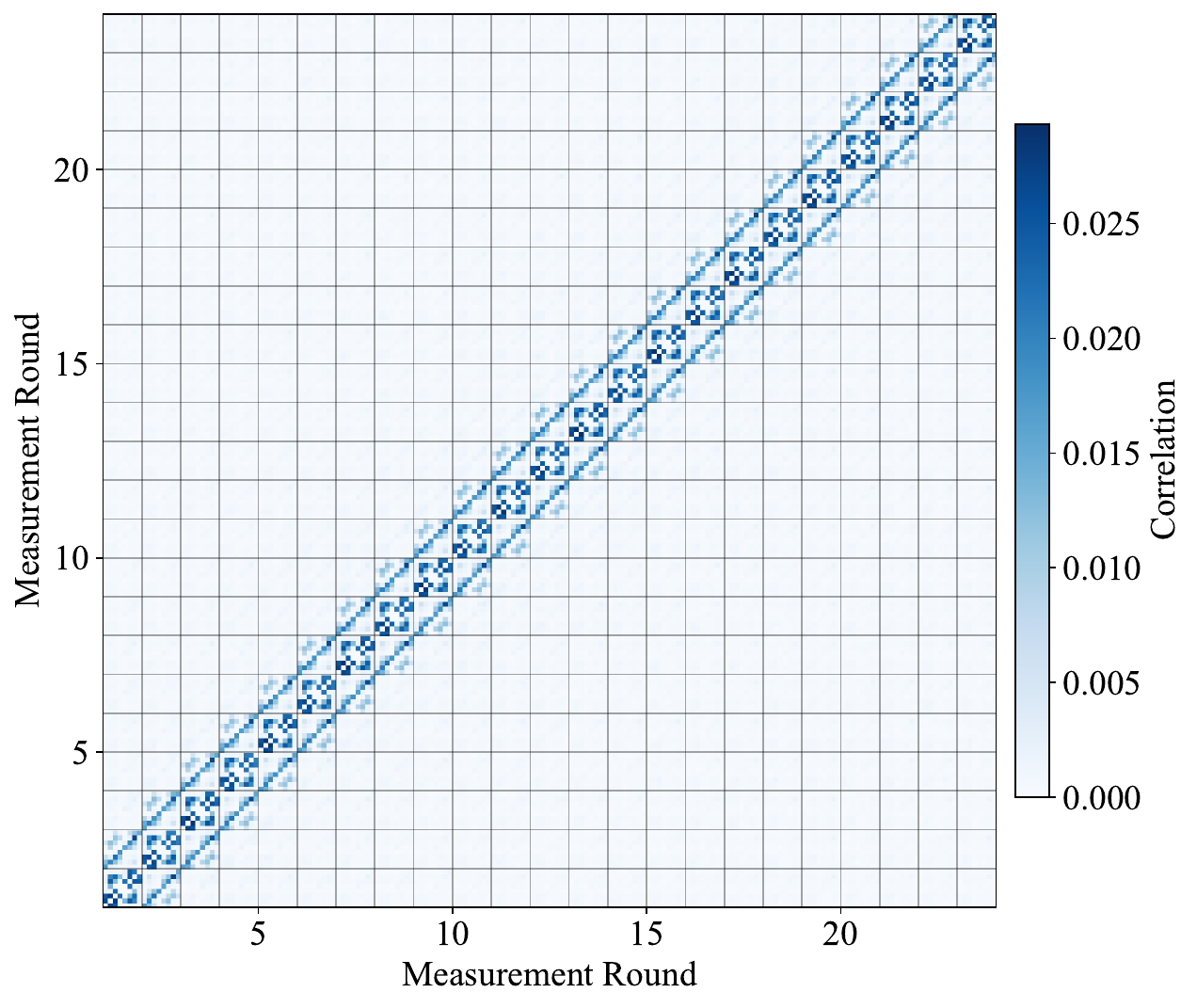}
    \caption{Correlation matrix for $d=3$.}
\end{subfigure}
\begin{subfigure}[]{0.5\linewidth}
    \centering
    \includegraphics[width=\linewidth]{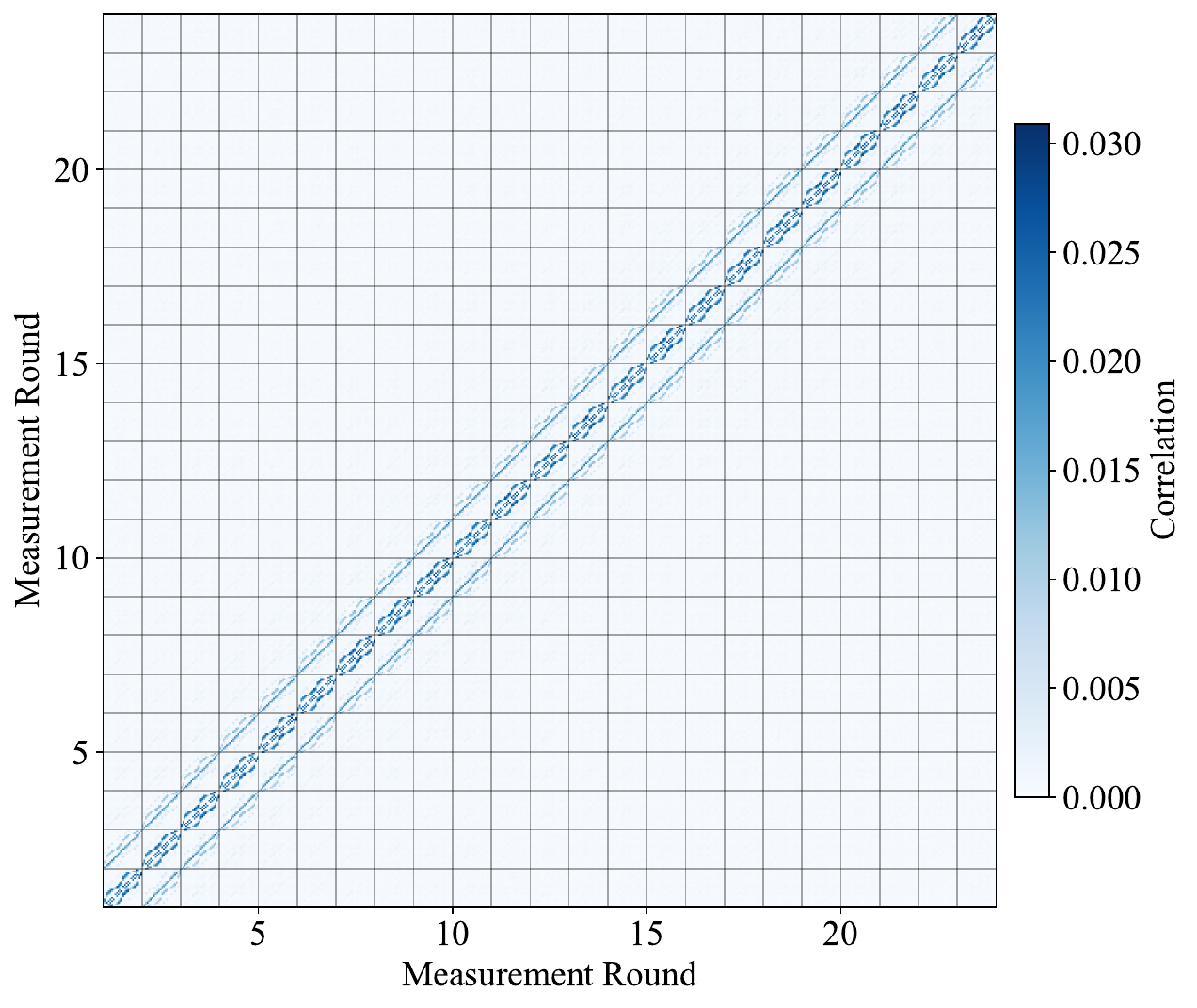}
    \caption{Correlation matrix for $d=5$.}
\end{subfigure}
\begin{subfigure}[]{0.5\linewidth}
    \centering
    \includegraphics[width=\linewidth]{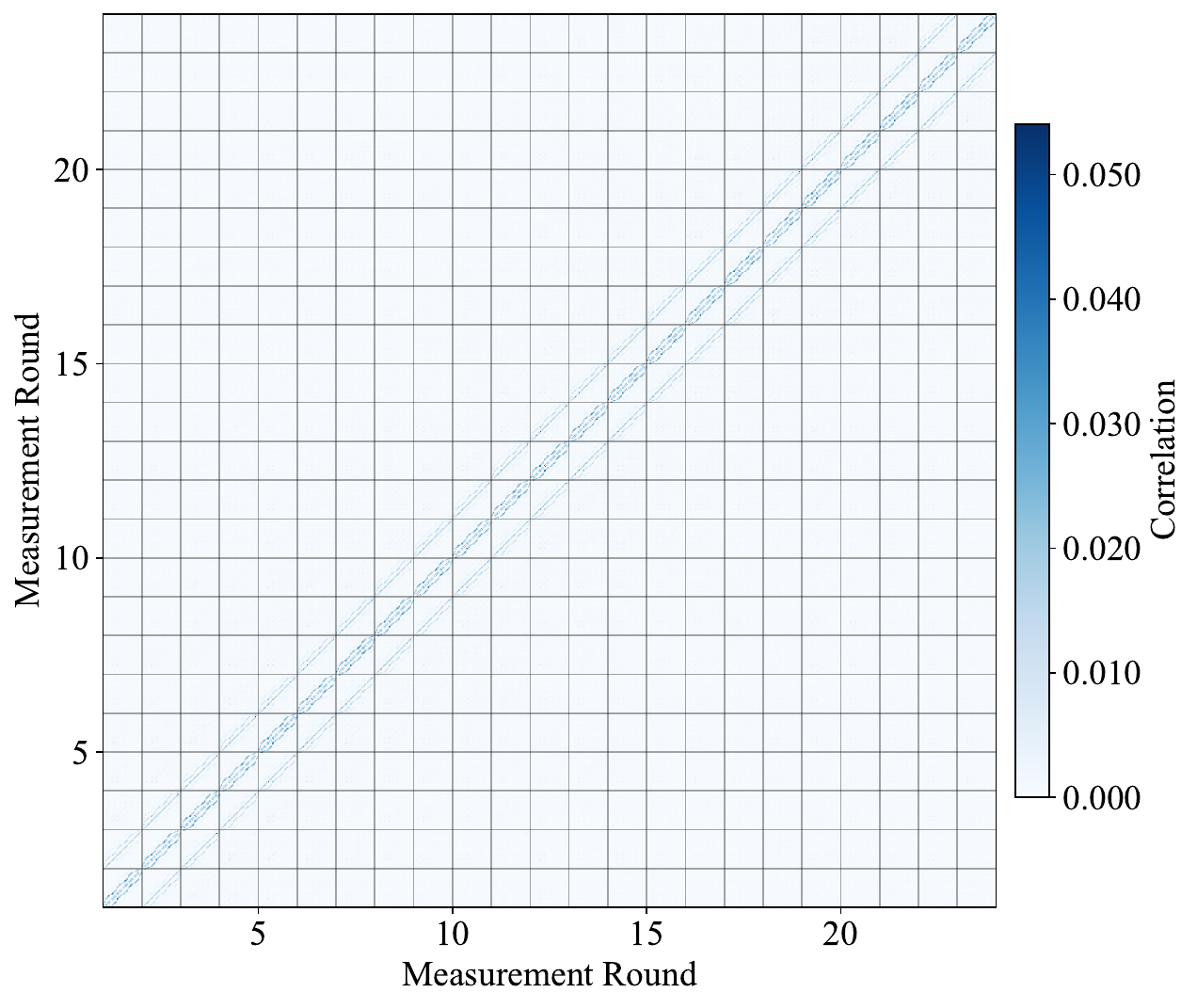}
    \caption{Correlation matrix for $d=7$.}
\end{subfigure}
\caption{Correlation visualization truncated to the first 25 rounds for clarity.}
\label{fig:correlation_matrix}
\end{figure}

We use detection event probability (DEP) as a compact model-agnostic probe of the effective noise experienced by the full QEC cycle.
DEP is especially informative for real devices because it can simultaneously reveal (i) temporal non-stationarity across repeated syndrome-extraction cycles (e.g., transient-to-steady-state behavior, end-of-sequence effects, or slow drift), and (ii) spatial heterogeneity across patch locations and detector types.
To expose these effects, we present: (a) DEP-versus-round trajectories for different patch instances and code distances, illustrating location-dependent noise levels and time-dependent variations, which is shown in Fig.~\ref{fig:def}; and (b) a round-round correlation matrix of detection events, which is shown in Fig.~\ref{fig:correlation_matrix}, where the off-diagonal structure directly indicates temporal correlations beyond independent-round noise.
Together, these visualizations provide direct evidence that the experimental noise contains much richer spatiotemporal structure than what is captured by the stochastic Pauli circuit-noise models used during large-scale pretraining.

\section{Theoretical Motivation for Parallel Decoding Without Merging}
\label{sec:appendix_decoding_without_merging}

It is difficult to mathematically analyze the performance of a neural decoder, but there is a relatively well-known proof that the MWPM decoder has a non-zero threshold for surface code memory experiments~\cite{fowler2012proof}. In this section, we will try to argue that this proof can probably be extended to a parallel decoder without merging when the size of the buffer region is $\ge d$. Though not directly about the neural decoder, this is indirect evidence that decoding without merging might be a viable option, and that consistency between windows is an essential ingredient. This also motivates the choice of the buffer size.

\subsection{Preliminaries}
In the context of a QEC procedure, we call an event $A$ \emph{exponentially unlikely} if there exists a threshold $p_\text{th} > 0$, such that
$$\Pr(A) = \text{poly}(d)\cdot \left(\frac{p}{p_\text{th}}^{\lceil d/2\rceil}\right).$$

Given a decoding graph $G = (V, E)$ and a subset $C \subseteq E$, we define the \emph{boundary} of $C$ as $\partial C = \{ v\in V \mid |\{e \in C \mid v \in e\}| \bmod 2 = 1\}$. We adopt the (slightly unusual but well-motivated) convention that ``virtual vertices'' of the entire decoding graph are not in $V$ (one can consider $G$ as a hypergraph with some edges with size $1$), and thus not in any $\partial C$, but ``virtual vertices'' on the time boundary of a window are in $V$ and thus can be in $\partial C$.

\medskip

We note that the parallel sliding-window decoder in~\cite{Tan2022ScalableSD} will probably \emph{not} work without merging if the inner decoder that decodes each individual window is allowed to connect the same two detection events across the seam with different paths in two adjacent windows. Therefore, for the following proof, we need an assumption about the inner decoder:

\begin{assumption}
\label{assumption:consistency}
    Let $G_1 = (V_1, E_1)$ and $G_2 = (V_2, E_2)$ be the decoding graphs of two adjacent windows with overlap, and let $C_1 \subseteq E_1$ and $C_2 \subseteq E_2$ be the corrections output by the inner decoder in these windows. For any $C'_1 \subseteq C_1$ and $C'_2 \subseteq C_2$, If $\partial C'_1 = \partial C'_2$, then $C'_1 = C'_2$. 
\end{assumption}

In plain words, if two subsets of corrections correct the same subset of detection events, then they are the same set of corrections. This property should hold for MWPM decoders if edge weights are perturbed to break ties.

\subsection{Proof}
\begin{theorem}
\label{thm:min_weight}
    When Assumption~\ref{assumption:consistency} holds, any physical error configuration that causes the parallel sliding-window decoder with the MWPM inner decoder to produce any non-trivial seam syndrome must have at least total weight $w_b/2$, where $w_b$ is the weighted buffer size (i.e., the shortest weighted distance on a window's decoding graph from a seam vertex to a virtual time boundary vertex).
\end{theorem}

Note that on a realistic decoding graph of the memory experiment, the weighted buffer size should be the buffer size times the weight of a vertical edge.

\begin{proof}[Proof]
Consider two adjacent windows $G_1 = (V_1, E_1)$ and $G_2 = (V_2, E_2)$ with window corrections $C_1$ and $C_2$ respectively. Consider the symmetric difference $D = C_1 \oplus C_2$. Since $\partial C_1$ and $\partial C_2$ must agree in the region where the two windows overlap (where they must both agree with the actual detection events observed), it follows that $\partial D = \partial C_1 \oplus \partial C_2$ must be entirely outside of this overlap region.

Suppose that the parallel sliding-window decoder did produce some non-trivial seam syndrome, and let $v$ be any one detection event in the seam syndrome. Then $D$ must touch $v$. Let $D'$ be the connected component of $D$ that includes $v$. Let $C'_1 = D' \cap C_1$ and $C'_2 = D' \cap C_2$. Obviously $C'_1 \ne C'_2$, so by Assumption~\ref{assumption:consistency}, we must have $\partial C'_1 \ne \partial C'_2$, i.e., $\partial D' \ne \emptyset$. However, as mentioned above, $\partial D' \subseteq \partial D$ must entirely fall outside the overlap region. Take any $u \in \partial D'$. The distance between $v$ (on the seam) and $u$ (outside the overlap region) must be at least the weighted buffer size, $w_b$.

We have proven that there must exist a path in $D$ between $v$ and $u$. The path has at least length $w_b$ and is composed of edges in $C_1$ and $C_2$, so either $w_{C_1} \ge w_b/2$ or $w_{C_2} \ge w_b/2$. Since the inner decoder is MWPM, the total weight of any correction must be no more than the total weight of the physical error configuration, so the latter must also be at least $w_b/2$.
\end{proof}

\subsection{Discussion}
We note that Theorem~\ref{thm:min_weight} may not be very useful in practice, since with constant physical error rate and $O(d^3)$ possible error locations, the total error weight in any window is almost certainly larger than $w_b$. What we really want to prove is:
\begin{conjecture}
\label{conj:fault_tolerance}
    When the buffer size is at least $d$, and Assumption~\ref{assumption:consistency} holds, it is exponentially unlikely that the parallel sliding-window decoder with the MWPM inner decoder produces any non-trivial seam syndrome.
\end{conjecture}

Ideally, we want to be able to use a counting argument like the one in~\cite{fowler2012proof}. However, this would require characterizing the locations of the actual physical errors (e.g., at least $m/2$ errors on a length-$m$ path), not just their total weight. An apparent difficulty is that, while the path in $D$ must reach $u$ from $v$ without touching a space boundary (if the path touches a space boundary on \emph{both} sides before exiting the overlap region, then both windows should decode the path in the same way), the actual physical errors can touch the space boundary. Even though the total weight of physical errors still must be at least equal to the total weight of corrections, there seems to be more freedom with the path it takes.

\subsection{Idea to prove Conjecture~\ref{conj:fault_tolerance}}
\begin{proof}[Proof]
We follow the proof of Theorem~\ref{thm:min_weight} up to the point where we get a path in $D$ between $v$ and $u$. Without loss of generality, we assume that $u$ is the only vertex on the path that falls outside of the overlap region. We denote that path as $P$, and its length (total weight) as $w_P$. We further denote $C_i \cap P$ as $C^P_i$. Then either $w_{C^P_1} \ge w_P/2$ or $w_{C^P_2} \ge w_P/2$. Without loss of generality, suppose that the former holds.

Below, we redefine the symbol $\partial$ so that it only includes \emph{real vertices} in the window $G_1$. This ensures that $\partial \mathcal{E} = \partial C_1$.

Now we find a subset $\mathcal{E}^Q$ of the physical errors $\mathcal{E}$ such that $\partial \mathcal{E}^Q = \partial C^Q_1$, for some $C^Q_1$ satisfying $C^P_1 \subseteq C^Q_1 \subseteq C_1$. We construct $\mathcal{E}^Q$ and $C^Q_1$ iteratively, starting from $\mathcal{E}^Q = \emptyset$ and $C^Q_1 = C^P_1$. Every step, as long as $\partial \mathcal{E}^Q \ne \partial C^Q_1$, we choose a vertex $t \in \partial \mathcal{E}^Q \oplus \partial C^Q_1$ arbitrarily (depending only on the current values of $\mathcal{E}^Q$ and $C^Q_1$). Since $\partial \mathcal{E} = \partial C_1$, we have that $t \in \partial (E - \mathcal{E}^Q) \oplus \partial (C_1 - C^Q_1)$. Therefore we can arbitrarily choose an edge incident to $t$ in either $E - \mathcal{E}^Q$ or $C_1 - C^Q_1$, and add it to $\mathcal{E}^Q$ or $C^Q_1$ respectively. This procedure must end because there are finitely many edges in $\mathcal{E}$ and $C_1$, and must result in $\partial \mathcal{E}^Q = \partial C^Q_1$.

Now observe that, \emph{a priori} (i.e., with $\mathcal{E}$, $\partial \mathcal{E}$, $C_i$... all unknown), all possible pairs $(\mathcal{E}^Q, C^Q_1)$ can be generated with a two-phase procedure.
{
\begin{itemize}
    \small
    \item First choose an arbitrary vertex $v$. Starting at $v$, every step move to an adjacent vertex, and choose whether the edge just traversed belongs to $C^P_1$ or $C^P_2$. Repeat this step until we reach a vertex $u$ outside of the overlap region, and we have the path $P$ as well as $C^P_1$.
    \item Then do the construction procedure in the previous paragraph, except that every step after determining $t$, we arbitrarily choose an edge incident to $t$ and arbitrarily choose either it belongs to $\mathcal{E}^Q$ or $C^Q_1$. When this procedure ends, we get a candidate $(\mathcal{E}^Q, C^Q_1)$.
\end{itemize}
}

Consider all such candidates where the whole procedure takes $n$ steps. Letting $w_\text{min}$ be the minimum weight of any edge in the decoding graph, and $n_1$ and $n_2$ be the number of steps each phase takes (thus $n_1 + n_2 = n$), we have:
$$w_P \ge n_1 \cdot w_\text{min}, \quad w_{C^P_1} \ge \frac{n_1}{2} \cdot w_\text{min},$$
$$w_{\mathcal{E}^Q} + w_{C^Q_1} \ge (\frac{n_1}{2} + n_2) \cdot w_\text{min}, \quad w_{\mathcal{E}^Q} \ge (\frac{n_1}{4} + \frac{n_2}{2}) \cdot w_\text{min} \ge \frac{n}{4} \cdot w_\text{min},$$
where the last step makes use of the fact that $C^Q_1$ is a min-weight correction for $\partial \mathcal{E}^Q = \partial C^Q_1$. The probability that all edges in the candidate $\mathcal{E}^Q$ are in the actual $\mathcal{E}$ is upper bounded by $\hat{p}_n = \exp \left(\frac{n}{4} \cdot w_\text{min}\right)$. Meanwhile, letting $k$ be the maximum degree of any vertex in the decoding graph, the number of all candidates is upper bounded by $\hat{N}_n = |V_\text{seam}| \cdot (2k)^{n}$. The total probability of those candidates are thus lower bounded by
$$\hat{p}_n \cdot \hat{N}_n = |V_\text{seam}| \cdot \left(2k\exp \frac{w_\text{min}}{4}\right) ^ n.$$

Therefore $\hat{p}_n \cdot \hat{N}_n$ is a geometric series, and for fixed $k$, when $w_\text{min}$ is large enough (i.e., when the physical error rate $p$ of the most likely error mechanism is low enough), this series decays exponentially with $n$. Furthermore, $n$ is lower bounded by the (unweighted) buffer size $b$\footnote{More accurately, $n_1$ is lower bounded by $b$ and $n_2$ is lower bounded by $\frac{w_b}{2w_\text{min}}$.}. By taking $b = 4\lceil d/2\rceil$, the leading term of this geometric series has the form
$$\hat{p}_b \cdot \hat{N}_b = |V_\text{seam}| \cdot \left((2k)^4\exp w_\text{min}\right) ^ {\lceil d/2\rceil} = \text{poly}(d)\cdot O(p)^{\lceil d/2\rceil}.$$
We can let $p$ be small enough that the common ratio of the geometric series is upper bounded by a constant (e.g., 1/2), so that the sum of this geometric series is still an exponentially unlikely probability.

\end{proof}

\begin{figure}
\centering
\includegraphics[width=\linewidth]{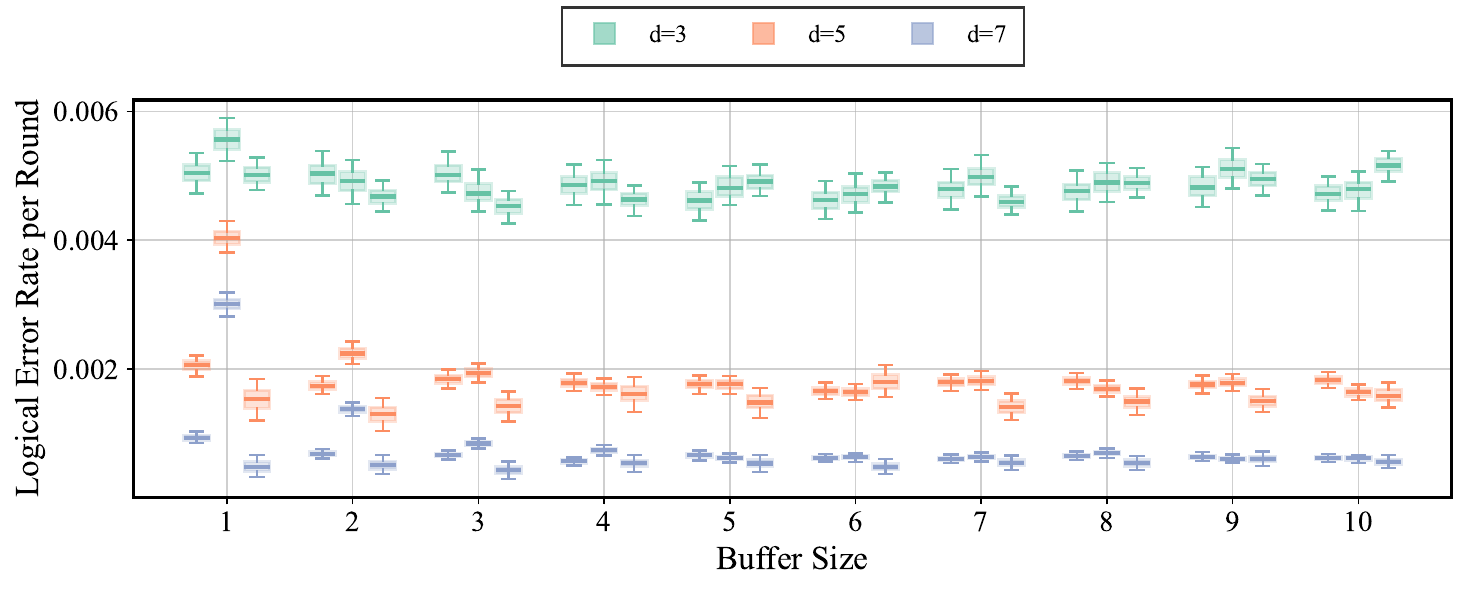}
\caption{Parallel window decoding without local merging. For memory experiments up to about $100$ syndrome measurement rounds under $p=0.003$, the decoding accuracy of parallel MWPM decoding with (left box) or without (middle box) merging converges with the global MWPM decoding (right box) as the buffer size approaches the code distance $d$.}
\label{fig:decode_without_merge}
\end{figure}

Aside from the mathematical proof, we also validate this observation experimentally, as shown in Fig.~\ref{fig:decode_without_merge}.

\end{appendices}

\newpage
\bibliography{reference} 
\end{document}